\documentclass[12pt]{article}
\pdfoutput=1
\usepackage{jcappub,natbib,float,caption,comment,bm}
\usepackage{graphicx,epsfig}
\usepackage[dvipsnames]{xcolor}
\usepackage[utf8]{inputenc}

\usepackage{slashed}

\newcommand{\fr}[2]{\mbox{$\frac{\,{#1}\,}{#2}$}}

\def\bga{\begin{aligned}}
\def\eda{\end{aligned}}
\newcommand{\beq}{\begin{equation}}
\newcommand{\eeq}{\end{equation}}
\newcommand{\bq}{\begin{equation}}
\newcommand{\eq}{\end{equation}}
\newcommand{\bay}{\begin{array}}
\newcommand{\eay}{\end{array}}
\newcommand{\beqa}{\begin{eqnarray}}
\newcommand{\eeqa}{\end{eqnarray}}
\newcommand{\beqs}{\begin{subequations}}
\newcommand{\eeqs}{\end{subequations}}

\def\nn{\nonumber}

\def\({\left(}
\def\){\right)}
\def\[{\left[}
\def\]{\right]}

\def\End{\end{document}}

\newcommand{\bea}{\begin{eqnarray}}
\newcommand{\eea}{\end{eqnarray}}

\def\dd{\text{d}}
\def\ii{\text{i}}
\def\GN{G_{\!N}^{}}
\def\MP{M_{\text{P}}^{}}
\def\leqq{\leqslant}
\def\geqq{\geqslant}
\def\To{\Rightarrow}

\newcommand\lsim{\mathrel{\rlap{\lower4pt\hbox{\hskip1pt$\sim$}}
        \raise1pt\hbox{$<$}}}
\newcommand\gsim{\mathrel{\rlap{\lower4pt\hbox{\hskip1pt$\sim$}}
        \raise1pt\hbox{$>$}}}

\title{Probing Dynamics of Boson Stars by Fast Radio Bursts and Gravitational Wave Detection
}
\author[a]{{\large Gongjun Choi,}}
\author[a,b,c,d]{{\large \,Hong-Jian He,}}
\author[b,a]{{\large \,Enrico D. Schiappacasse}}
\affiliation[a\,]{{\small Tsung-Dao Lee Institute, Shanghai 200240, China}}
\affiliation[b\,]{{\small School of Physics and Astronomy, Shanghai Jiao Tong University, Shanghai 200240, China}}
\affiliation[c\,]{{\small Institute of Modern Physics, Tsinghua University, Beijing 100084, China}}
\affiliation[d\,]{{\small Center for High Energy Physics, Peking University, Beijing 100871, China}}

\emailAdd{gongjun.choi@gmail.com}
\emailAdd{hjhe@sjtu.edu.cn}
\emailAdd{Enrico.Schiappacasse@sjtu.edu.cn}

\abstract{
\\
Boson stars may consist of a new type of light singlet scalar particles with
nontrivial self-interactions, and may compose a fraction of the dark matter in the Universe.
In this work, we study the dynamics of boson stars with Liouville and logarithmic scalar
self-interaction potentials as benchmarks. We perform a numerical analysis as well as
a semi-analytic study on how the compactness and the total mass will deviate from that of
the usual boson stars formed with a quartic repulsive self-interaction. We apply the
recently suggested Swampland conjecture to examine whether boson stars with such benchmark potentials
belong to the Landscape of a quantum gravity. Using the mass constraint on the macroscopic
compact halo object (MACHO) and the cold dark matter (CDM) isocurvature mode constraint from
the cosmic microwave background (CMB), we derive the allowed mass range of scalar particles
which compose the boson star. We further analyze applications of the lensing of fast radio bursts (FRBs)
and the gravitational wave (GW) detection to probe the presence of such boson stars and constrain
the parameter space of their corresponding models. We discuss how the two types of boson star potentials
can be discriminated by the FRB and GW measurements.
\\[3mm]
{e-Print: arXiv:1906.02094\,[astro-ph.CO]
and JCAP (2019), in Press. }
}

\begin{document}

\maketitle
%\flushbottom

\setcounter{page}{2}
\vspace*{4mm}
\section{Introduction}
\label{sec:section1}
\vspace*{1mm}

The Higgs boson discovery in 2012 has provided encouraging evidence that the spin- 0 scalar particles are among the most fundamental ingredients in nature, and there may well exist new singlet scalar particles beyond the standard model (SM) which are weakly coupled to the SM and may play a wide range of important roles, including the candidates for dark matter, dynamical dark energy, inflaton, axion, and some other things.

\vspace*{1mm}

If there exits a new scalar $\Phi$ in a dark sector (serving as a SM singlet), it may
couple to the SM sector very weakly via gravitational interactions. As the temperature of the Universe decreases in its history,
it is probable that such bosonic particles
become cold enough to sit
in the ground state at a certain moment.
This happens when the de Broglie wavelength of the bosons, $\lambda_{\rm dB}\sim m_{\Phi}^{-1/2}T^{-1/2}$,
becomes comparable to the interparticle distance between the particles.
Then, as a self-gravitating system, the clustering of Bose-Einstein condensate (BEC)
would tend to continue until gravitational attraction is balanced  by the repulsive quantum pressure due to the Heisenberg uncertainty principle. In consequence, a stable compact object (called boson star)
could form, and this may possibly provide a certain fraction of the dark matter in the Universe.

\vspace*{1mm}

With this motivation, there have been several works studying the dynamics of boson stars, including the possible maximum mass and compactness
of a boson star as determined by distinctive forms of the scalar potential. Given the competition
between gravitational attraction and repulsive quantum pressure, a possible scalar self-interaction
new force will cause a different hydrostatic equilibrium point, and thereby the characteristic physical
quantities describing the boson star will vary. As a simplest setup,
the stable boson star with a free massive scalar was studied in \cite{Kaup:1968zz}, and was investigated further \cite{Ruffini:1969qy} by using field quantization
of a real scalar field. This boson star (called mini-boson star) has a mass
$\sim\! m_{\rm P}^{2}/m_{\Phi}$, still less than the Chandrasekhar mass limit $\sim\!\! m_{\rm P}^{3}/m^{2}$\,
which is the maximum achievable mass for a fermion star. (Here $m_{\rm P}=G^{-1/2}_{{\rm N}}$ is the Planck mass.) Then, the scalar model with a repulsive quartic self-interaction was examined \cite{Colpi:1986ye} in the context of boson stars, and it was shown that the
total mass comparable to the Chandrasekhar limit can be realized. In the strong coupling regime, the compactness of this type of boson stars
was found to be as high as $\,C_{\max}^{}\!\simeq 0.16$\, \cite{AmaroSeoane:2010qx,Chavanis:2011cz}. Based on these works, many different potentials were further considered in the literatures\,\cite{Ho:1999hs,Mielke:1980sa,Friedberg:1986tp,Schunck:2003kk}.

\vspace*{1mm}

In this work, we will study the dynamics of boson stars with certain distinctive scalar self-interaction potentials. Regarding the new forces governing the hydrostatic equilibrium for the boson star other than the gravity and the quantum pressure, a nontrivial question is how the compactness and total mass would change, provided a infinite series of repulsive self-interaction forces are introduced.
Namely, we wonder how much increase from $\,C_{\max}^{}\!\simeq 0.16$\, can be observed
when the quantum pressure is put together with the repulsive force of an infinite series of
scalar self-interactions in competing with the gravitational attraction.
Would it be so significant that it approaches a value as high as the compactness
of a black hole?
If not, then how much does the infinite series of repulsive forces make
the stable boson stars
be different from the usual boson stars (including the mini-boson star or the boson star from
a repulsive quartic self-interaction)?
In addition, the same question can be asked for a scalar potential of which the expansion gives
infinite series of paired attractive and repulsive forces by having alternating signs. Answering these questions is valuable for identifying the source of a compact object-related
astrophysical signal that is characterized by its compactness and mass.
For instance, such astrophysical signals include gravitational waves (GWs) caused
by the merger of two binary compact objects or the fast radio bursts (FRBs)
due to the presence of a compact object serving as the lensing source. We further derive the allowed mass range of the scalar particles (which compose the
boson star) by using
(i) the current constraints on the fraction of the dark matter contributed by MACHO and the mass of MACHO,
and (ii) the current CMB constraint on CDM isocurvature modes.
(Here MACHO stands for the macroscopic compact halo object, and CDM for the cold dark matter.)
We study whether the FRB lensing and GW detection can be used to probe the currently allowed
scalar mass range. For this purpose, we will examine equilibrium configurations of the boson star
due to two benchmark potentials of scalar self-interactions --- the Liouville and Logarithmic potentials.

\vspace*{1mm}

This paper is organized as follows. In Sec.\,\ref{BSmodel}, we introduce a formalism for studying the dynamics of boson stars and set up two benchmark scalar potentials. In Sec.\,\ref{sec:results}, we present the results
of both numerical and analytic computation for the compactness and mass of boson stars. Then, we apply the Swampland conjecture to the results and check whether the boson star models under
consideration belong to the Landscape of a quantum gravity theory. In Sec.\,\ref{sec:AP}, we derive the allowed mass range of the scalar particles for forming a boson star, by using the current constraints on the MACHO mass and the fraction of the CDM from the MACHO,
and the current CMB constraint on CDM isocurvature modes. Then, we apply the two astrophysical measurements to probe the presence
of such boson stars and the parameter space of the benchmark scalar potentials. We study how the two types of boson star potentials can be discriminated
by the FRB and GW measurements. Throughout this paper, we will adopt the natural unit $c=\hbar=1$.
We denote the Planck mass by $\,m_{\rm P}=G_N^{-1/2}$ and the reduced Planck mass by $M_{\rm P}=(8\pi G_N^{})^{-1/2}$, where $G_N^{}$ is the Newton gravitational constant.
For the cosmological parameters, we use values based on Planck TT,TE,EE+lowE+lensing
at the 68\% confidence level in Ref.\,\cite{Aghanim:2018eyx}.

\vspace*{2mm}
\section{Boson Star Modeling}
\label{BSmodel}
\label{sec:2}

In this section, we present the full set of equations which governs the boson star solutions.
These equations include the Einstein equations for the spacetime geometry and the Klein-Gordon
equation of the scalar field. The equations of such a coupled system are usually called
Einstein-Klein-Gordon (EKG) equations (cf. Refs.\cite{Schunck:2003kk, Liebling:2012fv} for reviews).
Then, we proceed to introduce two self-interacting scalar potentials of different underlying physics.
We will motivate the choice of the two potentials. Finally, we discuss the definitions
of physical quantities characterizing a boson star, including the total mass and compactness.

\vspace*{2mm}
\subsection{Einstein-Klein Gordon Equations}
\label{sec:2.1}
\vspace*{1.5mm}

Boson star is a self-gravitating system comprised of particles which
correspond to a complex scalar field obeying the EKG equations.
We start by considering the following action,
\beq
S \,= \int\!\!\dd^4x\,\sqrt{-g\,} \left(\frac{R}{\,16\pi G_N^{}\,}
      + \mathcal{L}_{M}^{} \!\right) \,,
\label{action}
\eeq
where $R$\, is the Ricci curvature scalar, $g=\det\!\(g_{\mu \nu}^{}\)$
is the determinant of the metric tensor,
and $\mathcal{L}_M^{}$ is the Lagrangian density of the scalar field. For a complex scalar field $\Phi(r,t)$, we write the corresponding Lagrangian
\beq
\mathcal{L}_M^{} = -g^{\mu\nu}\partial_{\mu}\Phi^*\partial_{\nu}\Phi - U(|\Phi|^2),
\label{LM}
\eeq
where we require the scalar potential $U(|\Phi|^2)$ to be a function of the modulus of the scalar field.\footnote{Here we have dropped the cosmological constant term since its effect is negligible for the current boson star study according to \cite{He:2017alg}.}
This feature of the potential is crucial for making the action invariant under a global $U(1)$ symmetry.
Variations of the action (\ref{action}) with respect to the metric tensor and the scalar field
lead to the following EKG evolution equations,
\beqs
\beqa
R_{\mu \nu}^{} - \fr{1}{2}g_{\mu \nu}^{}R  &=&  8\pi G_{\!N}^{}T_{\mu \nu}^{}\,,
\label{EE}
\\[2mm]
\square \Phi - \frac{\dd V}{\dd |\Phi|^2} \Phi &=& 0\,,
 \label{eom}
\eeqa
\eeqs
where $\,\square$\, is the covariant D\textsc{\char13}\,Alembert operator,
$R_{\mu \nu}^{}\,$ is the Ricci tensor, and $\,T_{\mu \nu}^{}$ is the scalar
energy-momentum tensor,
\beqa
T_{\mu \nu} \,=\, \left( \partial_{\mu}^{}\Phi^{*} \partial_{\nu}^{}\Phi
+  \partial_{\mu}^{}\Phi\partial_{\nu}^{}\Phi^{*} \right)
-g_{\mu\nu}^{}\left[ \partial^{\alpha}\Phi^{*} \partial_{\alpha}^{}\Phi + U(|\Phi|^2) \right]\,.
\label{stress}
\eeqa
Equation (\ref{EE}) can be further expressed as $G_{\mu \nu}^{}\!=8 \pi\GN T_{\mu\nu}^{}$,\,
with $\,G_{\mu \nu}^{}\!= R_{\mu \nu}^{}\!-\fr{1}{2}g_{\mu \nu}^{}R$\, being the Einstein tensor.
The invariance of the action (\ref{action}) under a phase transformation,
$\,\Phi \to \exp(i\theta)\Phi$,\, implies a conserved Noether current
$J^{\mu} \!=\! \ii g^{\mu \nu}( \Phi^* \partial_{\nu}^{}\Phi - \Phi \partial_{\nu}^{}\Phi^*)$.\,
Thus, the corresponding Noether charge defines a conserved total number
of particles\,\cite{Ruffini:1969qy},
\beqa
N = \int\!\!\dd^3x\sqrt{-g\,}\, J^0 \,.
\label{Nparticles}
\eeqa
Stable general relativistic boson stars satisfy the relation
$\,M < Nm_{\Phi}^{}$\, due to the existence of a negative binding energy
associated with the spacetime geometry.

\vspace*{1mm}

The choice of a complex scalar field over a real scalar field has a reason.
As Friedberg, Lee and Pang showed\,\cite{Friedberg:1986tp},
a localized time-independent matter configuration does not exist for a real scalar field. Since the energy-momentum tensor (\ref{stress}) depends on the modulus of the field, the product
of gradients of the field and its conjugate, it is convenient to consider a harmonic ansatz
for the scalar field which ensures a time-independent gravitational field.
Thus, we write
\beqa
\Phi({\bf{r}},t) \,=\, \phi(r)\, e^{\ii\omega t}\,,
\label{ansatz}
\eeqa
%
%\hspace*{-2mm}
where $\,\phi \in \mathbb{R}$\, is the real radial profile of the scalar field
and $\,\omega\,$ a real angular frequency. We impose a spherically symmetric ansatz
to look for equilibrium configurations which correspond to minimum energy solutions.
With this ansatz, the energy-momentum tensor becomes time-independent, leading to static metric
functions. Thus, we can write
\beqa
\dd s^2 = -e^{\gamma(r)}\dd t^2 + e^{\lambda(r)}\dd r^2
+ r^2\!\left[\dd\theta^2\! + \sin^2(\theta)\dd\varphi^2\right]\!,
\label{metric}
\eeqa
%
%\hspace*{-2.5mm}
where $\,\gamma\!=\!\gamma(r)$\, and $\,\lambda\!=\!\lambda(r)$\,
depend only on the Schwarzchild-type radial coordinate $r$\,.\,
By substituting the ansatz (\ref{ansatz}) into Eqs.\eqref{eom}-\eqref{stress},
we obtain the temporal and radial components of the energy-momentum tensor and the equation of motion
of the scalar,
\beqs
\beqa
T_{t}^{t} &=& -U(\phi^{2})-\omega^2 e^{-\gamma} \phi^{2} -  e^{-\lambda}\phi'^{2} \,,
\label{Ttt}
\\[1mm]
T_{r}^r &=& -U(\phi^{2}) + \omega^2 e^{-\gamma} \phi^{2} + e^{-\lambda}\phi'^{2} \,,
\label{Trr}
\\
\phi'' &=& -\left[\frac{2}{\,r\,}+\frac{\,\gamma' \!-\! \lambda'\,}{2}\right]\!\phi'
+ e^{\lambda}\phi\!\left[\! \frac{\,\dd U(\phi^2)}{\dd\phi^2} - e^{-\gamma}\omega^2 \right]\!,
\\[-6mm]
\nn
\label{eom2}
\eeqa
\eeqs
%
%\hspace*{-2mm}
where a prime means a derivative with respect to $r$\,.\,
After computing $G^t_t$ and $G^r_r$, the first two components of the Einstein equations are given by
\beqs
\beqa
e^{-\lambda}\!\(\frac{\lambda'}{r} + \frac{e^{\lambda}}{r^2}-\frac{1}{r^2} \)
&=\,& 8\pi \GN \!\!\left[ U(\phi^{2}) + \omega^2 e^{-\gamma}\phi^{2} + e^{-\lambda}\phi'^{2}  \right]\!,
\label{EE1}
\\
e^{-\lambda}\!\(\frac{\gamma'}{r} - \frac{e^{\lambda}}{r^2}+\frac{1}{r^2}\)
&=\,& 8\pi\GN \!\!\left[ -U(\phi^{2}) + \omega^2 e^{-\gamma}\phi^{2} + e^{-\lambda}\phi'^{2}  \right]\!.
\label{EE2}
\eeqa
\eeqs

For the later computational convenience, we introduce the following dimensionless variables,
$\tilde{r}=r m_{\Phi}^{}$,\, $\tilde{\omega}=\omega/m_{\Phi}^{}$,\,
and $\tilde{\phi}^{2}=8\pi\GN \phi^{2}$.\,
These rescaled variables lead to the resultant coupled system of differential equations,
\beqs
\beqa
\gamma'(\tilde{r}) &=&
 \frac{\,e^{\lambda(\tilde{r})} \!-\! 1\,}{\tilde{r}} + \tilde{r}e^{\lambda(\tilde{r})}\!\!\left[-\tilde{U} + \tilde{\omega}^2e^{-\gamma(\tilde{r})}\tilde{\phi}^{2} + \tilde{\phi}'^{2}e^{-\lambda(\tilde{r})}  \right]\!,
\label{EQ1}
\\
\lambda'(\tilde{r}) &=&
 -\left[\frac{\,e^{\lambda(\tilde{r})} \!-\! 1\,}{\tilde{r}}\right]
 + \tilde{r}e^{\lambda(\tilde{r})}\!\!\left[\tilde{U} + \tilde{\omega}^2e^{-\gamma(\tilde{r})}\tilde{\phi}^{2} + \tilde{\phi}'^{2}e^{-\lambda(\tilde{r})}  \right]\!,
\label{EQ2}
\\
\tilde{\phi}''(\tilde{r}) &=&
-\left[ \frac{2}{\tilde{r}} + \frac{\,\gamma'(\tilde{r})\!-\!\lambda'(\tilde{r})\,}{2}
\right]\!\tilde{\phi}' + e^{\lambda(\tilde{r})}\tilde{\phi}\!
\left[ \frac{\dd\tilde{U}}{\dd\tilde{\phi}^2} -e^{-\gamma(\tilde{r})}
\tilde{\omega}^2\right]\!,
\label{EQ3}
\eeqa
%\eeqs
%
%\hspace*{-2.5mm}
where the prime denotes a derivative with respect to $\tilde{r}$\,.\,
In the above, we have defined dimensionless quantities for convenience,
\beqa
\,\tilde{U} \!= (8\pi\GN / m_{\Phi}^2) U(\phi^{2}),\quad \,\phi^2 = \tilde{\phi}^2/8\pi\GN\,.
\eeqa
\eeqs

\vspace*{1mm}
\subsection{Scalar Self-Interaction Potentials}
\label{sec:potential}
\vspace*{1.5mm}

Different from the mini-boson star without any self-interaction\,\cite{Kaup:1968zz},
it was shown\,\cite{Colpi:1986ye} that the maximum total mass of a fermion star
(Chandrasekhar mass limit) could be mimicked by a boson star provided the scalar field
is allowed to have repulsive quartic self-interaction with a suitably chosen particle mass
and interaction strength.
The underlying physics is that the repulsive force due to quartic self-interaction with positive coupling ($\lambda_{\Phi}> 0$) together with the quantum pressure of boson particles can better compete with the attractive gravitational force, leading to an equilibrium radius of a larger size.\footnote{In addition to the repulsive quartic self-interaction,
the back reaction from curvature is another source of the repulsive force against gravitational attraction as discussed before \cite{Croon:2018ybs}.} Hence, along this line of thinking, it is natural to wonder what else can achieve the higher compactness.
In other words, we may ask whether compactness of a boson star increases significantly
under the presence of additional sources of either of a repulsive or attractive force.
As a first possibility, there may be a series of repulsive self-interaction terms.
To avoid a divergent repulsive force, we demand that the added new repulsive higher order
self-interaction terms becomes smaller as the expansion order increases.
In this case, can we have a stable boson star? And if so, will there be a significant change
in compactness as compared to the mini-boson star? To address these questions, we may consider the following modified $U(1)$ Liouville potential\,\cite{Schunck:1999zu},
\beqa
U_{\text{Liouville}}^{}(|\Phi|^2)
\,=\, f^2 m_{\Phi}^2\!\(\text{e}^{\frac{|\Phi|^2}{f^2}} \!-1 \)\!,
\label{VLiouville}
\eeqa
where $f$\, serves as a coupling strength parameter. (For the boson star studies in the literature, the scalar $\Phi$ is normally defined as a SM singlet and joins only the gravitational interactions in addition to its self-interactions. The above effective nonlinear scalar self-interactions may be induced by the quantum gravity effects around the Planck scale.) For $|\Phi|\!<f$\, and using Eq.(\ref{ansatz}), we may expand the potential around $\,\phi = 0\,$
as follows,
\beqa
U_{\text{Liouville}}^{} \,=\, m_{\Phi}^2\phi^2 + \frac{m_{\Phi}^2\phi^4}{2 f^2}
+ \frac{m_{\Phi}^2\phi^6}{6 f^4} + \frac{m_{\Phi}^2\phi^8}{24 f^6}
+ \mathcal{O}\!\(m_{\Phi}^2\phi^{10}\!/\!f^8\)\,,
\label{TaylorLiouvill}
\eeqa
%
%\hspace*{-2mm}
where the scalar mass term arises from the leading order of the expansion. We have modified the usual Liouville potential to respect the global $U(1)$ symmetry
which ensures the particle number conservation. Indeed, the above modified Liouville potential
has the desired property of generating an infinite series of repulsive interaction terms.
On the other hand, such a convex exponential potential of a scalar field has been applied to
a variety of problems in cosmology, ranging from dark matter\,\cite{Malakolkalami:2016thy},
inflation\,\cite{Taylor:2000ze,Hawking:1998ub,Tsujikawa:2004dm,Handley:2014bqa},
to dark energy modeling\,\cite{Amendola:1999er,Brax:1999gp}.

\vspace*{1mm}

As a second option, we may consider a potential of which the higher order contributions
largely cancel with each other pairwise due to different signs,
but the net effect is not negligible and differs from a single quartic self-interaction.
Note that this case can have two possibilities:
the sign of the quartic coupling can be either of positive
($\lambda_\Phi^{}\!>0$) or negative ($\lambda_\Phi^{}\!<0$).
For $\lambda_\Phi^{}\!>0$, the net force due to scalar self-interactions is repulsive and
the resulting compactness turns out to exceed that of the mini-boson star ($\simeq\!0.08$)
as shown in \cite{Croon:2018ybs} with the cosine potential (cf.\ its Fig.10).
For the current study, we will examine the other possibility with a {\it negative} quartic coupling,
in which case the net force due to self-interactions should be attractive.
A well-known example of this kind is the QCD axion potential (cf. Refs.\cite{Schiappacasse:2017ham,Hertzberg:2018lmt,Hertzberg:2018zte,Visinelli:2017ooc,Chavanis:2017loo} for reviews of axion stars).
For this purpose, we consider the following logarithmic scalar potential,
\beqa
U_{\text{Log}}^{}(|\Phi|^2) \,=\,
f^{2}m_{\Phi}^2\, \text{log}\!\left(\frac{\,|\Phi|^2}{f^2} +1\right)\!.
\label{VLog}
\eeqa
%
%\hspace*{-2mm}
For $|\Phi|<f$, we make Talyor-expansion of the scalar potential around $\,\phi = 0$\,
and use Eq.\,(\ref{ansatz}) to derive the following form,
\beqa
U_{\text{Log}}^{} \,=\,
m_{\Phi}^2\phi^2 - \frac{\,m_{\Phi}^2\phi^4}{2f^2} + \frac{\,m_{\Phi}^2\phi^6}{3f^4}
- \frac{\,m_{\Phi}^2\phi^8}{4f^6} + \mathcal{O}\!\(m_{\Phi}^2\phi^{10}\!/\!f^8\) ,
\eeqa
%
%\hspace*{-2mm}
where the mass term arises from the leading order expansion again.
Note that both potentials are functions of the modulus of the scalar field as required
by the global $U(1)$ symmetry. This logarithmic type potential can appear as the effective potential
for a flat direction of the scalar field in the gauge mediated supersymmetry breaking
scenario\,\cite{deGouvea:1997afu}. Besides, thermal logarithmic potential can be induced
due to interaction of the real scalar field with other fields in the primordial plasma
after inflation for a supersymmetric theory\,\cite{Anisimov:2000wx,Mukaida:2012qn}.
Motivated by these, the logarithmic potential of the same form as Eq.(\ref{VLog})
for a real scalar field was studied in the context of I-ball formation in \cite{Kasuya:2002zs,Kawasaki:2013hka}.

\vspace*{1mm}

To make the computation simpler, we define a dimensionless parameter $\,\tilde{f}$\, by
$\,\tilde{f}^2=8\pi\GN f^2$.\, We also define a dimensionless coupling strength $\,\Lambda \!=\! 1/\tilde{f}^2$\,.\,
Note that, for the special case of a usual quartic self-interaction, $U_{\text{quartic}}^{} \!= m_{\Phi}^2\Phi^2 + \fr{1}{2}\lambda_{\Phi}^{}\Phi^4$,\,
we have $\,\Lambda =  \lambda_{\Phi}^{}M_{\text{P}}^2/m_{\Phi}^2$.\footnote{For the case of the quartic self-interaction, this definition for $\Lambda$ agrees with the definition given in Ref.~\cite{Colpi:1986ye} after a field and strength coupling redefinition. Our Lagrangian in Eq.~(\ref{LM}) becomes equal to the Lagrangian in that reference by making $\Phi \rightarrow \bar{\Phi}/\sqrt{2}$ and $\lambda_{\Phi} \rightarrow 2\bar{\lambda}_{\Phi}$. Thus, $\Lambda =  \bar{\lambda}_{\Phi} / (4\pi G_N m^2_{\Phi})$ as Eq.(2) of~\cite{Colpi:1986ye}.} \,
Then, in terms of the dimensionless variables defined above, we can convert the potentials
into the following form,
\beqs
\beqa
U_{\text{Liouville}}(\phi^2) \!\times\! (8\pi\GN /m^2_{\Phi})
&\,=\,& \tilde{f}^2\!\(\! \text{e}^\frac{\,\tilde{\phi}^2}{\tilde{f}^2}\!-1 \!\)
\equiv\, \tilde{U}_{\text{Liouville}} \,,
\hspace*{10mm}
\\[1mm]
U_{\text{Log}}(\phi^2) \!\times\! (8\pi\GN /m^2_{\Phi})
&\,=\,& \tilde{f}^2\,\text{log}\!\(\! \frac{\,\tilde{\phi}^2}{\tilde{f}^2} + 1 \!\!\)
\equiv\, \tilde{U}_{\text{Log}}\,.
\hspace*{10mm}
\eeqa
\eeqs
In the above, the dimensionless potentials
$\tilde{U}_{\text{Liuoville}}^{}$ and $\tilde{U}_{\text{Log}}^{}$
are the same as what appeared in Eqs.\eqref{EQ1}-\eqref{EQ3}.

\vspace*{2mm}
\subsection{Mass and Compactness of the Boson Star}
\label{sec:quantities}
\vspace*{1.5mm}

For a boson star, the physical quantities of interest are its total mass and compactness.
The total mass can be computed as, $\,M\!= m(r\!\to\! \infty)$, where
$(\dd m/\dd r) \!=\! - 4\pi r^2 T_t^t$.\, Using the dimensionless variables defined in the previous subsection,
we derive the following
\beqa
\tilde{m}(\tilde{r}) \,=\, \frac{1}{2}\int_{0}^{\tilde{r}}\!
\left[ \tilde{U} + \tilde{\omega}^2e^{-\gamma}\tilde{\phi}^2+e^{-\lambda}
\!\!\(\!\frac{\dd\tilde{\phi}}{\dd\tilde{r}'} \!\)^{\!\!\!2}  \right]\!\tilde{r}'^2 d\tilde{r}' \,,
\label{massformula}
\eeqa
where the dimensionless mass parameter is $\tilde{m}(\tilde{r})=(m_{\Phi}/m^2_{\text{P}})m(r)$.\,
Since we are working in an asymptotically flat spacetime, $\,\tilde{M} \!=\! \tilde{m}(\tilde{r}\!\to\! \infty)$\, in
Eq.(\ref{massformula}) agrees with the ADM-mass,
which is obtained from equalizing the radial component
of the metric \eqref{metric} with the radial component of the Schwarzschild metric,
\beqa
\tilde{m}(\tilde{r})\!\left|_{\tilde{r}\to\infty}^{}
\,=\, \left[\(1-e^{-\lambda(\tilde{r})}\)\frac{\tilde{r}}{2}\right]
\right|_{\tilde{r}\rightarrow\infty}^{} .
\label{ADM}
\eeqa

The conserved total number of particles associated with a boson star is calculated from Eq.\eqref{Nparticles}.
In terms of the dimensionless variables defined earlier, we have
\beqa
\tilde{N} = \int_0^{\tilde{r}}\!\! d\tilde{r}\, \tilde{r}^2
\tilde{\omega}\,\tilde{\phi}^2\,\text{e}^{\frac{\lambda-\gamma}{2}}\,,
\eeqa
where $\,\tilde{N}\! = (m^2_{\Phi}/m^2_{\text{P}})N$\, and
$\,\tilde{r}\!\to\! \infty$\, are understood.

\vspace*{1mm}

Since the radial profile of scalar field vanishes at the physical infinity, the scalar field has non-compact support. This means that boson stars do not have a hard surface and thus the definition
of a radius $R$ would be ambiguous. This ambiguity is also transferred
to the definition of compactness.
Following Ref.\,\cite{AmaroSeoane:2010qx},
we define an effective compactness,
\beqa
C(\phi_0^{},f) \,=\, \frac{\,0.99 M(\phi_0^{},f)\GN\,}{R_{99}^{}(\phi_0^{},f)}\,,
\label{Cdef}
\eeqa
where $R_{99}^{}$ refers to the radius at which the $99\%$ of the boson star mass $M$ is enclosed.
We will see that the effective compactness is a function of
the central scalar field value $\phi_0^{}$ and the coupling strength parameter $f$ of the theory.

\vspace*{2mm}
\section{Analysis and Results}
\label{sec:results}
\label{sec:3}
\vspace*{1.5mm}

In this section, we first present a numerical approach to solve the Einstein-Klein-Gordon equations in Sec.\,\ref{sec:numerical}. Then, we apply the recently suggested Swampland criteria to examine whether boson stars with the two scalar potentials of Sec.\,\ref{sec:swampland} belong to the Landscape of a quantum gravity theory. Finally, we present a semi-analytic approach in Sec.\,\ref{semianalytic} to estimate the maximum total mass, the minimal radius and maximum compactness of boson stars.

\vspace*{2mm}
\subsection{Numerical Implementation and Results}
\label{sec:numerical}
\label{sec:3.1}
\vspace*{1.5mm}

The equilibrium configurations associated with boson stars
are found by solving numerically the coupled system of Eqs.\eqref{EQ1}-\eqref{EQ3}
under suitable boundary conditions: (i).~$\gamma(\tilde{r}\!\to\!\infty) = \lambda(\tilde{r}\!\to\!\infty)
      = \tilde{\phi}(\tilde{r}\!\to\!\infty)=0$\,
ensures asymptotic flatness;
(ii).~$\lambda(\tilde{r}=0)=\tilde{\phi}'(\tilde{r}=0)=0$, $\gamma(\tilde{r}=0)=\gamma_0$, and $\tilde{\phi}(\tilde{r}=0)=\tilde{\phi}_0$ ensure regularity at the center.
For a given initial central value of the scalar field, $\tilde{\phi}_0^{}$,
the whole problem is simply reduced to an eigenvalue problem for the angular frequency $\tilde{\omega}$\,.\,
The value $\gamma_0^{}$ can be arbitrarily chosen before numerical computation.
We can always rescale the time variable in Eq.(\ref{metric})
to satisfy the asymptotic flatness $\gamma(\tilde{r}\!\to\!\infty)=0$.\,
For a radius greater than $2\GN M$,\, the metric becomes the Schwarzschild metric.

\vspace*{1mm}

There exists a discrete spectrum of eigenfrequencies $\tilde{\omega}$ for each value
$\tilde\phi_0^{}$. The lowest eigenfrequency $\tilde{\omega}_0^{}$
corresponds to the ground state configuration, whose radial profile $\phi(\tilde{r})$
does not include nodes. The other eigenfrequencies correspond to excited state configurations
with zeroes in their radial profile. As mentioned above, we are mainly interested in
the ground state scalar field configuration for boson stars. Since the excited states
would decay to the ground state by emission of scalar and gravitational
radiations\,\cite{Ferrell:1989kz}, they are not so relevant for our purpose.

\vspace*{1mm}

We execute a numerical integration of the coupled equations of Eqs.\eqref{EQ1}-\eqref{EQ3}
by using the Mathematica Software. The upper limit of the integration,
$\tilde{r}_{\max}^{}$, is chosen to be much greater than the characteristic radius
of the profile of the scalar field. According to the presence or absence of zeroes
in the radius profile $\tilde{\phi}(\tilde{r})$, we define a range of eigenfrequencies
$[\tilde{\omega}_a^{}, \tilde{\omega}_b^{}]$ such that
$\tilde{\omega}_a^{} \lesssim \tilde{\omega}_0^{} \lesssim \tilde{\omega}_b^{}$.
After that, we perform a search via the bisection method around the unknown
$\tilde{\omega}_0^{}$
based on the way in which $\tilde{\phi}(\tilde{r})$ diverges.
Near $\tilde{\omega}_0^{}$ (or any eigenfrequency), the numerical solution diverges
upward (downward) if the value chosen for $\omega$ is smaller (larger) than
$\tilde{\omega}_0^{}$\,.\,
Once we obtain a profile which does not diverges in the spatial region
$[0, \tilde{r}_{\max}^{}]$,\, we continue to apply the binary search until
${|\tilde{\phi}(\tilde{r}_{\max}^{})|} < \epsilon$,\,
where $\epsilon$ is the desired tolerance.
Finally, we impose asymptotic flatness on the temporal component of the metric
by rescaling the value of $\,\gamma_0^{}\,$ and $\,\tilde{\omega}_0^{}\,$
as $\gamma_0^{}  \!\to\! \gamma_0^{} \!+\!{\ln}(a)$ and
$\tilde{\omega}_0^{} \!\to\!  \tilde{\omega}_0^{}\sqrt{a}$,
where
$\ln(a) \equiv - [\gamma(\tilde{r}_{\max}^{})\!+\!\lambda(\tilde{r}_{\max}^{})]$.

\vspace*{1mm}

Depending on the scalar potential of the boson star model, the binary search
of the lowest eigenfrequency could be very challenging. For instance, a high level of numeric precision is required in the case of
ground state solutions for solitonic boson stars. These solutions turn out to be extremely sensitive to the tiny changes of the eigenfrequency value as a consequence
of its steep radial profile\,\cite{Macedo:2013jja}.
For the boson star models of our current interest,
the level of complexity of numerical calculations is similar to that for the mini-boson stars or massive stars with a quartic self-interaction. A detailed analysis of numerical computations for boson stars
was given before \cite{Lai:2004fw}.

\vspace*{1mm}

\begin{figure}[t]
\centering
\includegraphics[scale=0.30]{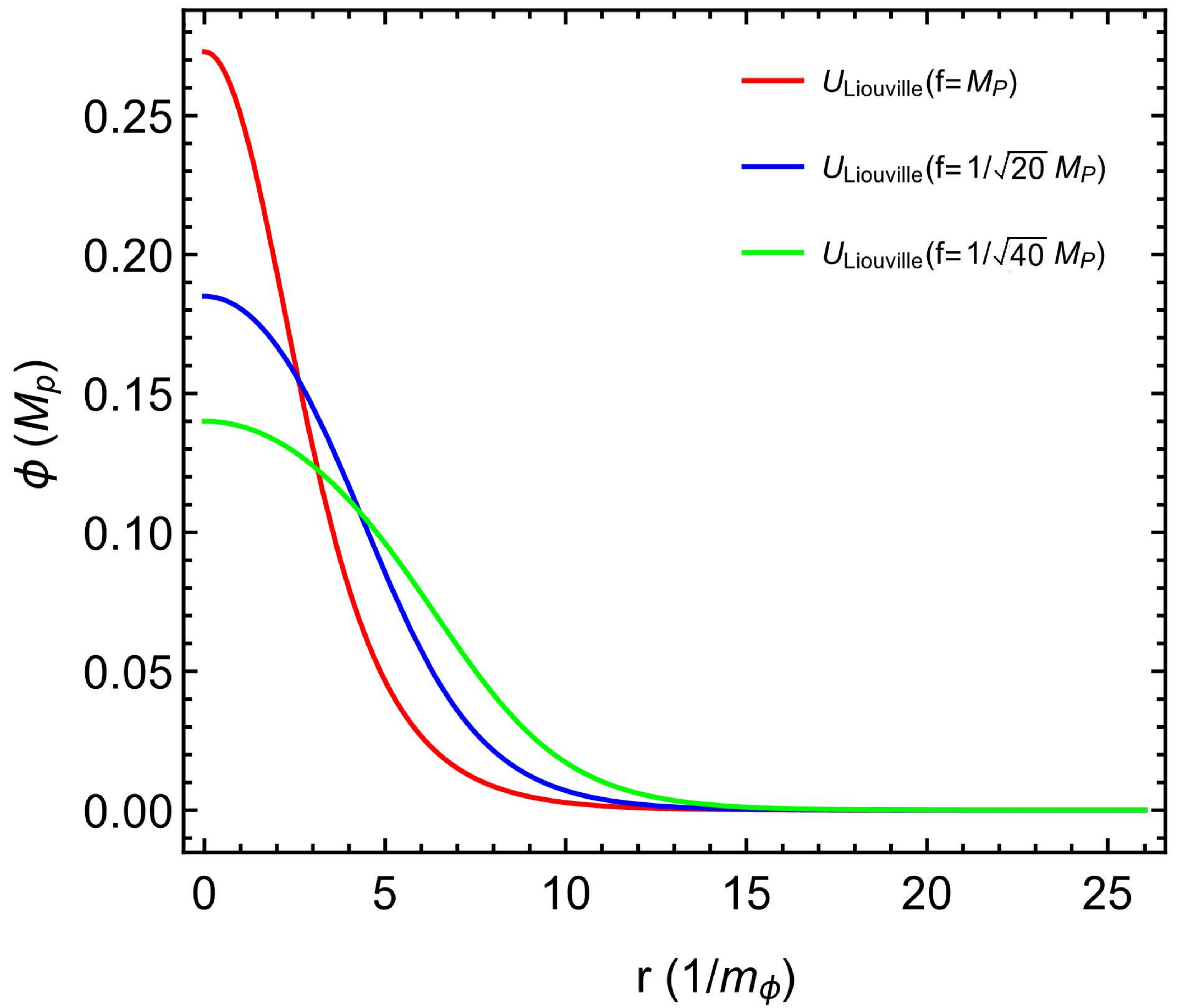}
\caption{Radial profile of the scalar field as a function of the radius for different values of $\phi_0$ and the coupling strength parameter,  in the case of $U(1)$ Liouville potential . The field  and the radius are shown in units of $M_{\text{p}}$ and $1/m_{\phi}$, respectively. The red, blue, and green curves correspond to the ground state solutions for $(0.273,M_\text{p})$,$\,(0.185,1/\sqrt{20}\, M_\text{p}\,)$,\, and $(0.140,1/\sqrt{40}\, M_\text{p})$ values of $(\phi_0,f)$, respectively. In particular,
the red curve is obtained under the suitable values $\,\omega = 0.84732123346818\,m_{\phi}^{}$ and $\gamma_0^{} = -0.80267626206664$\,.}
\label{RProfile}
\label{fig:1}
\vspace*{4mm}
\end{figure}
\begin{figure} %[h]
\centering
\hspace*{-5.4mm}
\includegraphics[scale=0.29]{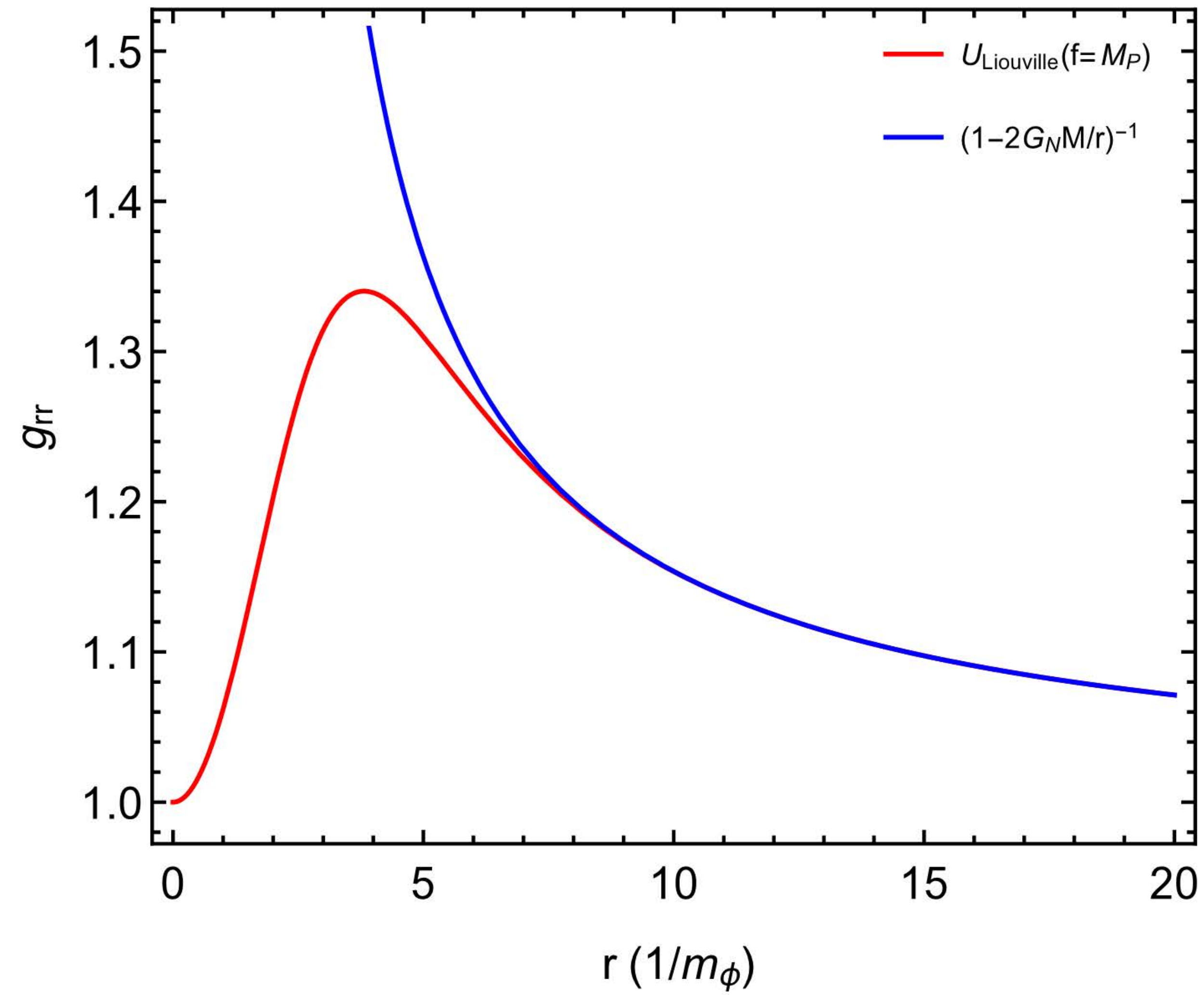}
\includegraphics[scale=0.29]{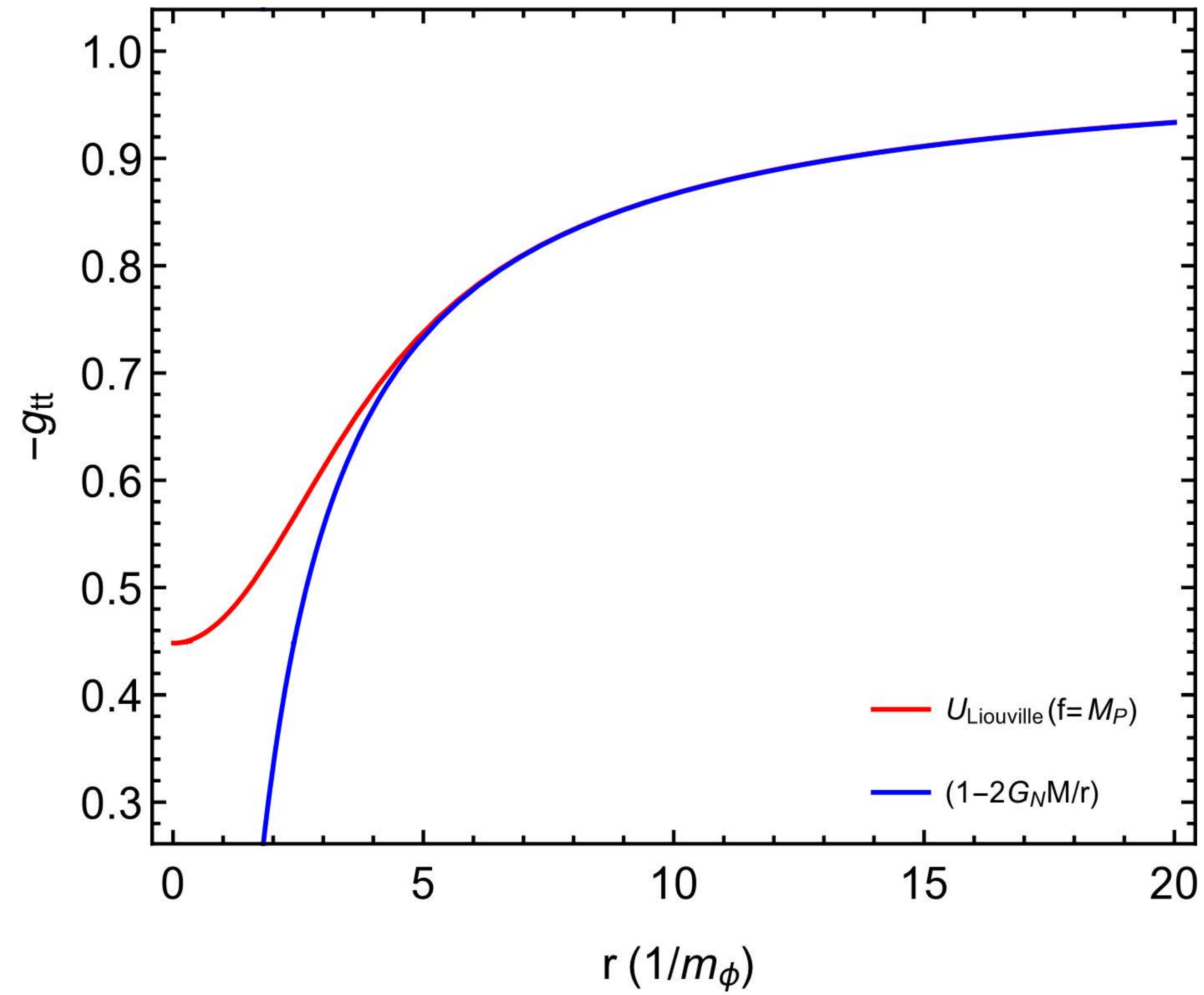}
\caption{Radial and temporal components of the metric (\ref{metric}) (red curves)
for the ground state solution of a boson star with the $U(1)$ Liouville potential
($f \!=\! M_{\text{P}}^{}$). This solution corresponds to the radial profile with $f \!=\! M_{\text{P}}^{}$ in Fig.\,\ref{RProfile}. The blue curves are the radial and temporal components of
the corresponding Schwarzschild metric. The radius is expressed in units of $1/m_{\Phi}$.}
\label{Metricsexamples}
\label{fig:2}
\end{figure}

Stability properties of boson stars were studied in the literature
both analytically\,\cite{Croon:2018ybs, Gleiser:1988ih, Lee:1988av, Jetzer:1989us}
and numerically\,\cite{Seidel:1990jh, Hawley:2000dt, Guzman:2004jw, Kusmartsev:1990cr}.
Mass and compactness of the ground state of a boson star depend on the central value
of the scalar field ($\phi_0^{}$) and the scalar potential $U(|\Phi|^2)$.
For the case of the mini-boson stars with free scalar field, there exists
a critical point for the central value of the scalar field, $\phi_0^{\star}$,
beyond which the ground state is unstable under small radial perturbations.
As $\phi_0^{}$ increases, stable configurations have larger masses but smaller
effective radius, leading to a greater compactness.
This behavior continues until the central value of the scalar field reaches $\phi_0^{\star}$,
where the total mass of the star encounters a turnaround.
This turnaround implies a maximum allowed mass for a boson star in the ground state,
which is found to be $\, M_{\max}^{}\!\!=\!0.633\,m^2_{\text{P}}/m_{\Phi}^{}$
\cite{Kaup:1968zz, Ruffini:1969qy}.
More generic potentials which include one or more self-interaction terms
added to the mass-term show the same stability features as the case of the free-field potential
\cite{Colpi:1986ye,Ho:1999hs,Mielke:1980sa,Schunck:1999zu}.
%\ref{Metricsexamples}

\vspace*{1mm}

Figure\,\ref{RProfile} shows the radial profile of the ground state of a boson star with a $U(1)$ Liouville potential for different values of $\phi_0$ and $f$.
As we expect for any boson star ground state, these radial profiles of the scalar field
do not contain any node and, starting from a central value $\phi_0^{}$,
go to zero as the radius increases. In particular, while the red curve in Fig.\,\ref{RProfile} shows the radial profile of the ground state solution for $\phi_0=0.273\,M_{\text{p}}$ and $f = M_{\text{p}}$, the red curves in Fig.\,\ref{Metricsexamples} show the radial and temporal components of the metric associated with this solution. From Fig.\,\ref{Metricsexamples}, we see that for the region outside the boson star, the metric components of Eq.(\ref{metric}) overlaps the corresponding Schwarzschild metric components (shown by the blue curves).

\begin{figure}[t]
\centering
\includegraphics[scale=0.4]{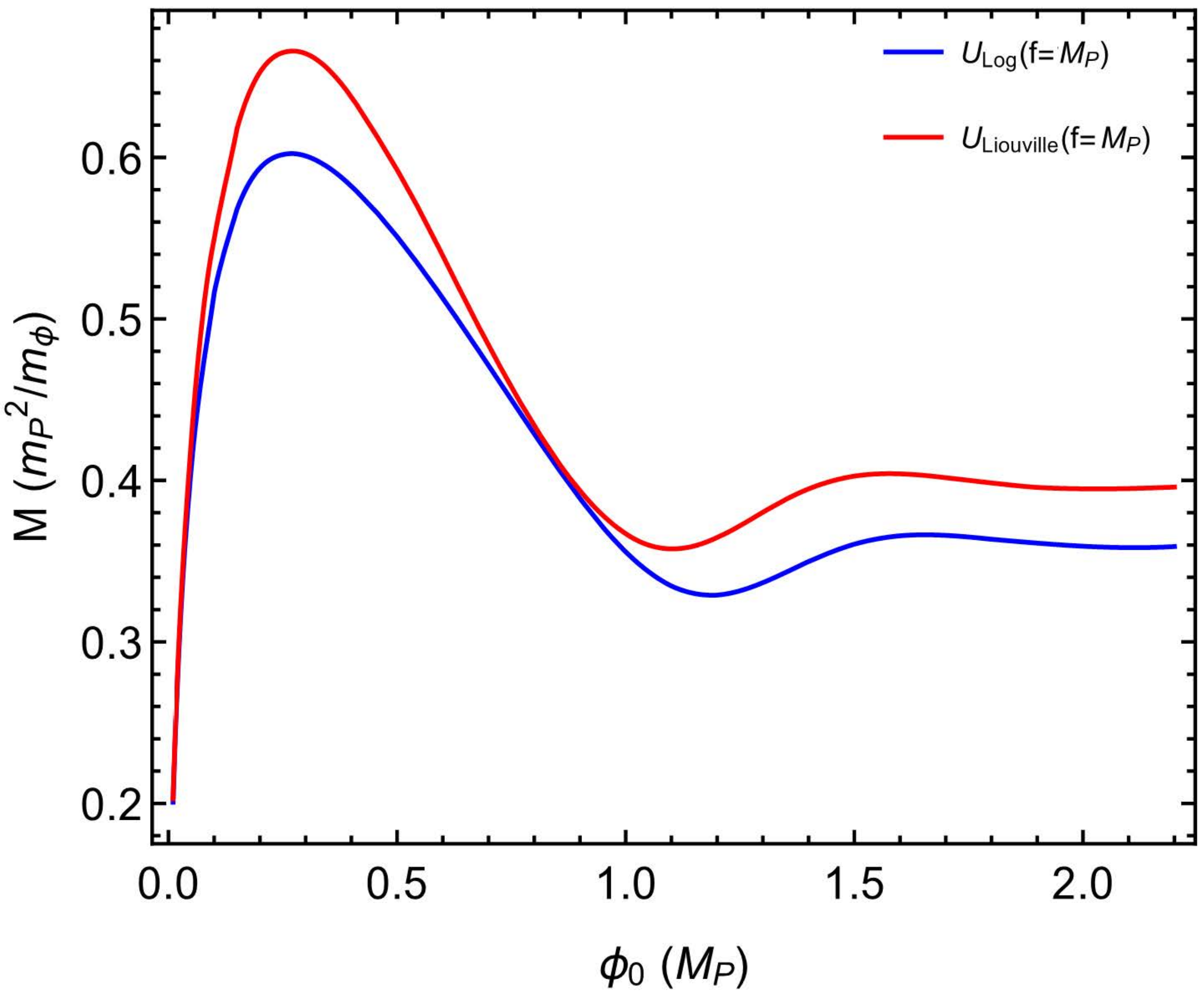}
\caption{Boson star mass $M$ (in units of $m^2_{\text{p}}/m_{\Phi}$) of the ground state configurations
as a function of the central value of the scalar field $\phi_0^{}$ (in units of $M_{\text{p}}$)
for the scalar potentials $U_{\text{Liouville}}^{}$ (red curve) and
$U_{\text{Log}}^{}$ (blue curve). In both cases, the coupling strength parameter $f$ is
set as $\,f \!=\! M_{\text{P}}$\,.}
\label{Massphi1}
\label{fig:3}
\end{figure}

\newpage

In Fig.\,\ref{Massphi1}, we present the boson stars mass $M$
as a function of the central value of the scalar field $\phi_0^{}$
for the scalar potentials $U_{\text{Liouville}}^{}$ (red curve) and $U_{\text{Log}}^{}$ (blue curve).
In both cases, we have set the coupling strength parameter to be $\,f = M_{\text{P}}^{}$.\,
We compute the mass of the ground state of these boson stars by using Eq.(\ref{massformula}),
which agrees with the ADM-mass in Eq.(\ref{ADM}).
The maximum mass for both potentials is obtained at the critical central value
of the field $\,\phi_0^{\star} \!\approx\! 0.27$\,.
The total mass of the star shows a turnaround at this critical central value.
All configurations {on the left-hand (right-hand) side
of the critical point} $\phi^{\star}_{0}$ are stable (unstable).
For the two scalar potentials $U_{\text{Liouville}}^{}$ and  $U_{\text{Log}}^{}$, we have $\,M_{{\max}}^{} \!= 0.666\,m^2_{\text{P}}/m_{\Phi}^{}$\, and
$\,M_{{\max}}^{}\! = 0.602\, m^2_{\text{P}}/m_{\Phi}^{}$,\, respectively.
In comparison to the maximum mass for a free-field potential,
the repulsive self-interaction terms present in the expansion of
the $U(1)$ Liouville potential enhance the value of $M_{{\max}}^{}$.\,
In contrast, the lower value of $\,M_{\max}^{}\,$ for the case of the
$U(1)$ logarithmic potential is caused by the presence of the net
effective attractive self-interactions from its expansion series. For the case of $U_{\text{Liouville}}^{}$ potential (as shown by the blue curve in Fig.\,\ref{Massphi1}), the value of $M_{{\max}}^{}$ and the $M(\phi_0^{})$ curve are
in full agreement with \cite{Schunck:1999zu}.

\vspace*{1mm}

\begin{figure}[t]
\centering
\hspace*{-3.5mm}
\includegraphics[width=7.5cm,height=5.8cm]{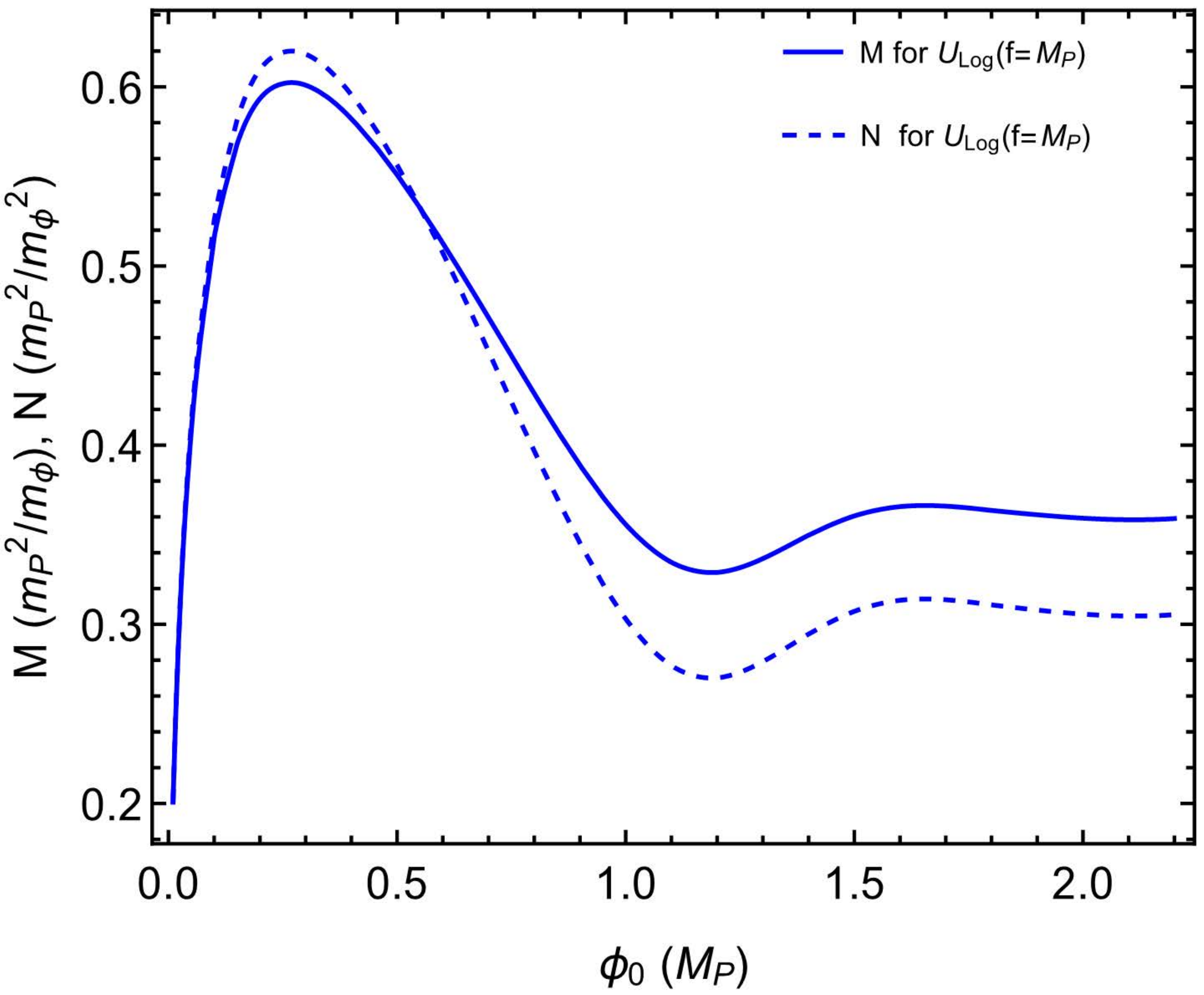}
\includegraphics[width=7.5cm,height=5.8cm]{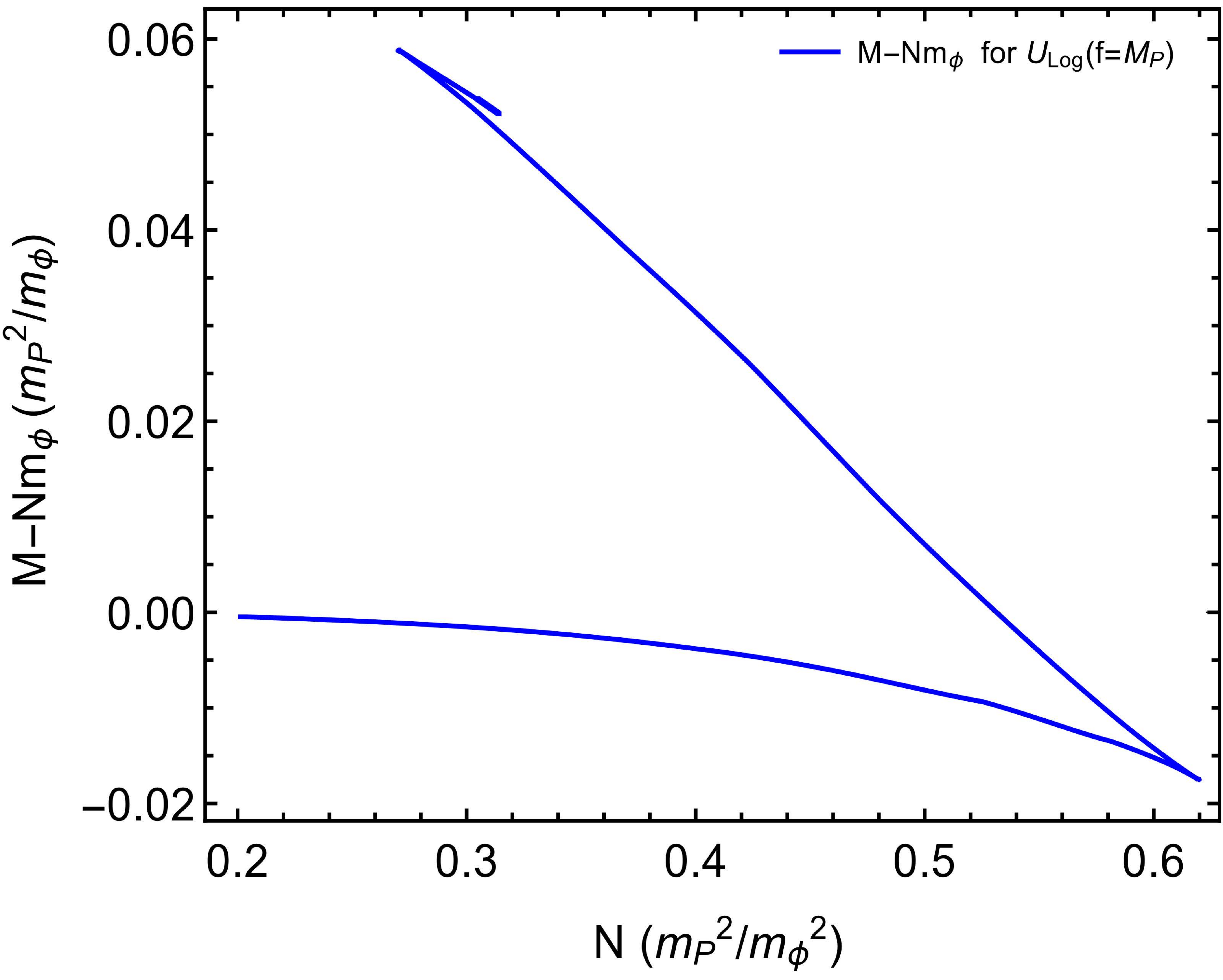}
\caption{Analysis of the ground state configurations for boson stars with a logarithmic
potential $U_{\text{Log}}^{}$, where the coupling strength parameter is $\,f = M_{\text{P}}^{}$.
The mass $M\,$ and number of particles $N\,$ of the boson star are shown as functions
of the central value of the scalar field $\phi_0^{}$ (left panel) and
the corresponding bifurcation diagram (right panel). $M$, $N$, and $\phi_0$ are shown in units of $m^2_{\text{p}}/m_{\Phi}$, $m^2_{\text{p}}/m^2_{\Phi}$, and $M_{\text{p}}$, respectively.  }
\label{Massphi}
\label{fig:4}
\end{figure}

In Fig.\,\ref{Massphi}, we further present the relation between the mass $M\,(\text{in units of}~m^2_{\text{p}}/m_{\Phi})$
and number of particles $N\,(\text{in units of}~m^2_{\text{p}}/m^2_{\Phi})$ of a boson star with the logarithmic potential $U_{\text{Log}}^{}$ (left panel)
and its respective bifurcation diagram (right panel).
The bifurcation diagram show cusps which denote changes in the stability properties
of the boson star configurations. The first branch is the only stable branch.
As we mentioned earlier, a negative binding energy,
$E_\text{B}^{}\! = M-Nm_{\Phi}^{}\!<0$\,,\,
is a necessary (but not sufficient) condition for the stability.

\vspace*{1mm}

\begin{figure}[t]
\centering
\hspace*{-5mm}
\includegraphics[width=7.4cm,height=5.6cm]{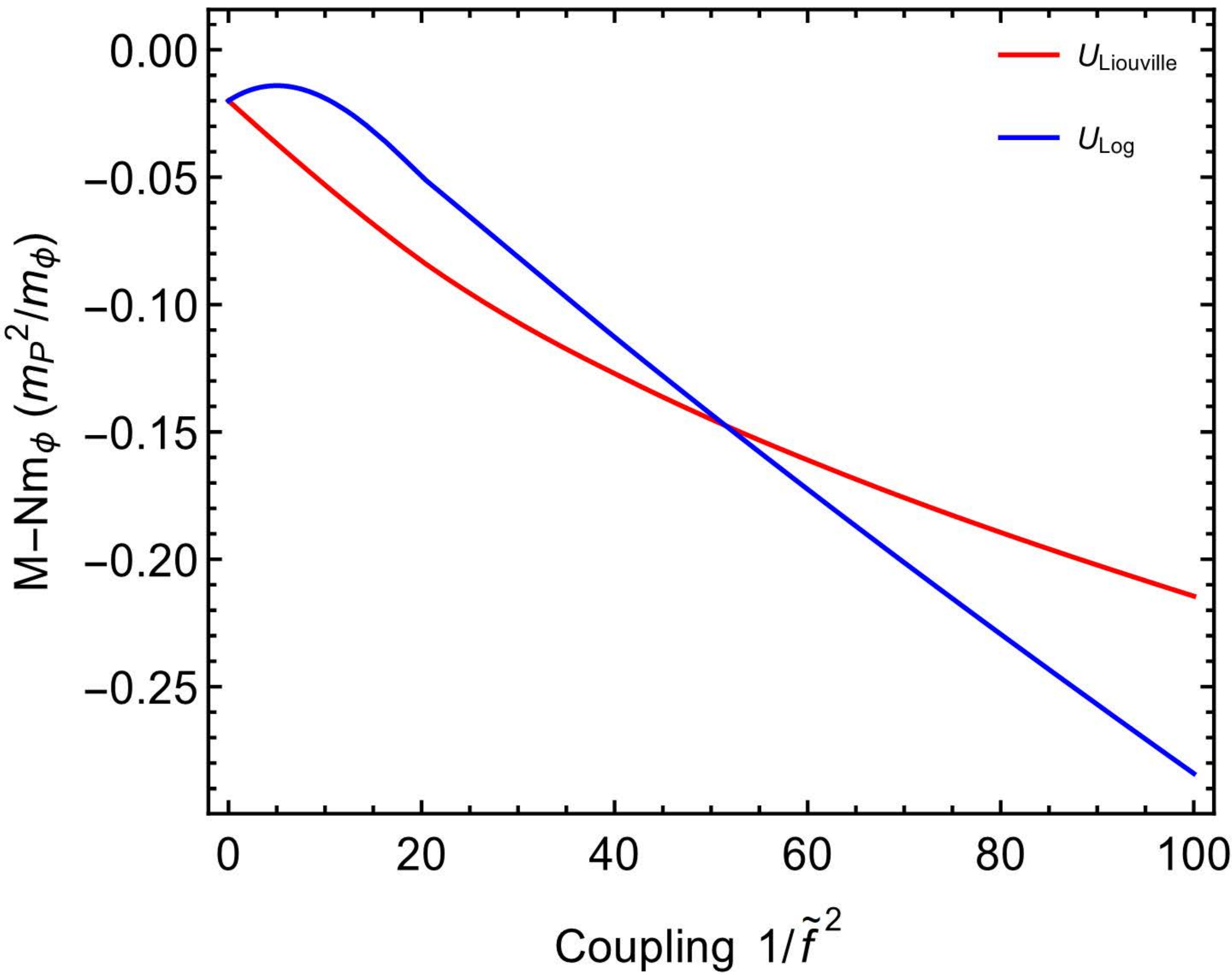}
\includegraphics[width=7.4cm,height=5.6cm]{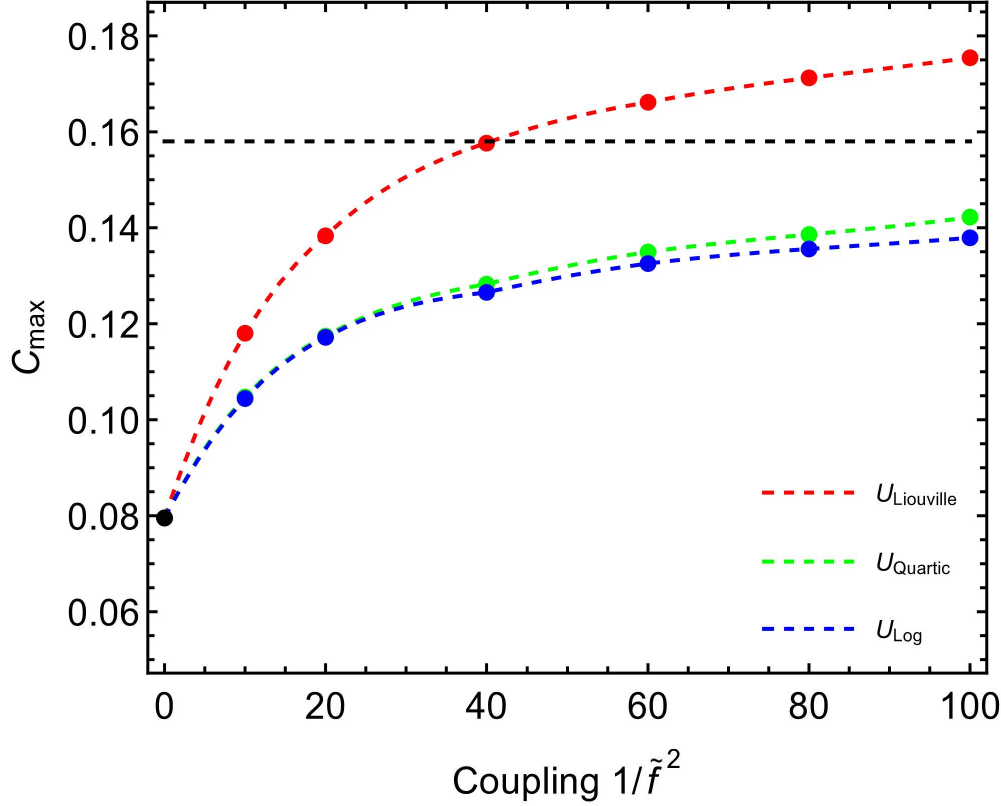}
\caption{Left Panel: Binding energy of the boson star,
$\,E_\text{B}^{}\!=\! M\!-\!Nm_{\phi}^{}\,$,\,
is shown as a function of the self-interaction coupling strength $\Lambda=1/\tilde{f}^{2}$ for scalar potentials
$U_{\text{Liouville}}^{}$ (red curve) and $U_{\text{Log}}^{}$ (blue curve).
Right Panel: Effective compactness $\,C_{\max}^{}\,$ is depicted as a function of
the self-interaction coupling strength for scalar potentials $U_{\text{Liouville}}^{}$ (red dashed curve)
and $U_{\text{Log}}^{}$ (blue dashed curve).
For comparison, we show the result for the usual repulsive quartic self-interaction potential,
$U_{\text{Quartic}}^{}$ (green dashed curve), including the asymptotic value of its effective
compactness when the coupling strength
$\,\Lambda \!=\! \lambda_{\Phi}^{}M_{\text{P}}^2/m_{\Phi}^2$
goes to infinity: $C^{\text{Quartic}}_{\text{max}}(\Lambda \!\to\! \infty)\simeq 0.158$
\cite{AmaroSeoane:2010qx} (black dashed line). The three curves converge to an initial point
in the weak coupling limit (black dot), corresponding to the compactness
of the ground state of a mini-boson star.}
\label{Compactness}
\label{fig:5}
\end{figure}

In the left panel of Fig.\,\ref{Compactness},
we present the negative binding energy of the boson star,
$\,E_\text{B}^{}\!=\! M\!-\!Nm_{\Phi}^{}\,$,\,
as a function of the self-interaction coupling strength $1/\tilde{f}^{2}$. For the Liouville potential $U_{\text{Liouville}}^{}$,
we observe that increasing self-interaction coupling strength raises the magnitude of the
negative binding energy $E_{{\rm B}}$. But, for the logarithmic potential $U_{\text{Log}}^{}$,
the magnitude of the binding energy first decreases up to the coupling strength
$\Lambda\!\sim\!5$ and then increases for $\Lambda\!\gtrsim\!5$.\,
We will examine the reason of the different behaviors of $E_{\rm B}^{}$ for the two scalar potentials in Sec.\,\ref{semianalytic}.

\vspace*{1mm}

As we will show in the next subsection, the compactness is one of the most relevant
physical quantities for studying the dynamics of boson stars
because of its connection to possible astrophysical signatures.
The right panel of Fig.\,\ref{Compactness} presents the evolution
of the maximum compactness reached for ground state configurations
of boson stars with respect to the coupling strength $\Lambda$. Here we have used the total mass $M$ of the boson star rather than $Nm_{\Phi}^{}$
for computing the maximum compactness $C_{\max}^{}$.
For comparison, we also show the case of the usual repulsive quartic self-interaction
potential $U_{\text{Quartic}}^{}$,\,
in addition to the potentials $U_{\text{Liouville}}^{}$ and  $U_{\text{Log}}^{}$.\,
As anticipated, we see that all three cases converge to the compactness of the mini-boson stars, $\,C_{\max}^{}\!\simeq 0.08$\, (shown as black dot), in the weak coupling limit.
However, their behaviors deviate from each other as the coupling strength increases.
The compactness for the potential $U_{\text{Log}}^{}$ increases with the coupling
strength, but is slightly smaller than that of the potential $U_{\text{Quartic}}$.\,
For instance, for $U_{\text{Log}}^{}$ and $\,f \!= 0.1M_{\text{P}}^{}$,\,
we find that its compactness is only less than that of $\,U_{\text{Quartic}}^{}$
by $0.3\%$.
The combination of an attractive leading order self-interaction term with alternating
repulsive and attractive higher order terms from expanding the $U(1)$ logarithmic potential,
produces a similar effect to the compactness of a potential with repulsive quartic
self-interaction alone.
However, the situation changes for the case of $U_{\text{Liouville}}^{}$.
The net effect of all repulsive self-interaction terms from its expansion series
increases the compactness significantly above that of the conventional
$U_{\text{Quartic}}^{}$ potential.
Note that for the $U_{\text{Liouville}}^{}$ potential with
$\,f \approx 1/\!\sqrt{40} \,M_{\text{P}}^{}$,\,
the compactness reaches the asymptotic value of that
for the $U_{\text{Quartic}}^{}$ potential,
$C^{\text{Quartic}}_{\max}\!(\Lambda \!\!\to\!\! \infty)\!\simeq\! 0.158$
\cite{AmaroSeoane:2010qx}. In the next subsections, we will use the compactness value
for both potentials with $f = 0.1M_{\text{P}}^{}$
as representatives of a strong coupling regime,
where $\,C_{\max}^{\text{Liouville}} \!\simeq 0.176$.\,
In Sec.\,\ref{semianalytic}, we will return to the analysis of compactness by using a semi-analytic approach.

\vspace*{2mm}
\subsection{Compactness in Landscape of Quantum Gravity}
\label{sec:swampland}
\label{sec:3.2}
\vspace*{2mm}

Although we motivated the scalar potentials introduced in Sec.\,\ref{sec:potential}, one may further wonder whether these consistent-looking scalar theories coupled to gravity could be UV-completed by a consistent quantum gravity theory. \textit{When the answer is positive, the effective field theories (EFT) are said to reside in the Landscape of a quantum gravity theory.} In contrast, if the answer is no, they are said to form \textit{the Swampland} according to \cite{Vafa:2005ui,Ooguri:2006in}. As a way to distinguish EFTs in the Landscape from those in the Swampland, certain criteria are suggested to be checked. Throughout these checks, we want to examine whether $C_{\max}^{}$ obtained in Sec.\,\ref{sec:numerical} is from the Landscape or the Swampland. The similar question was discussed in \cite{Herdeiro:2018hfp} for the cases of
the mini-boson stars and of boson stars with a positive quartic self-interaction potential.

\vspace*{1mm}

\begin{figure}[b]
\centering
\includegraphics[scale=0.43]{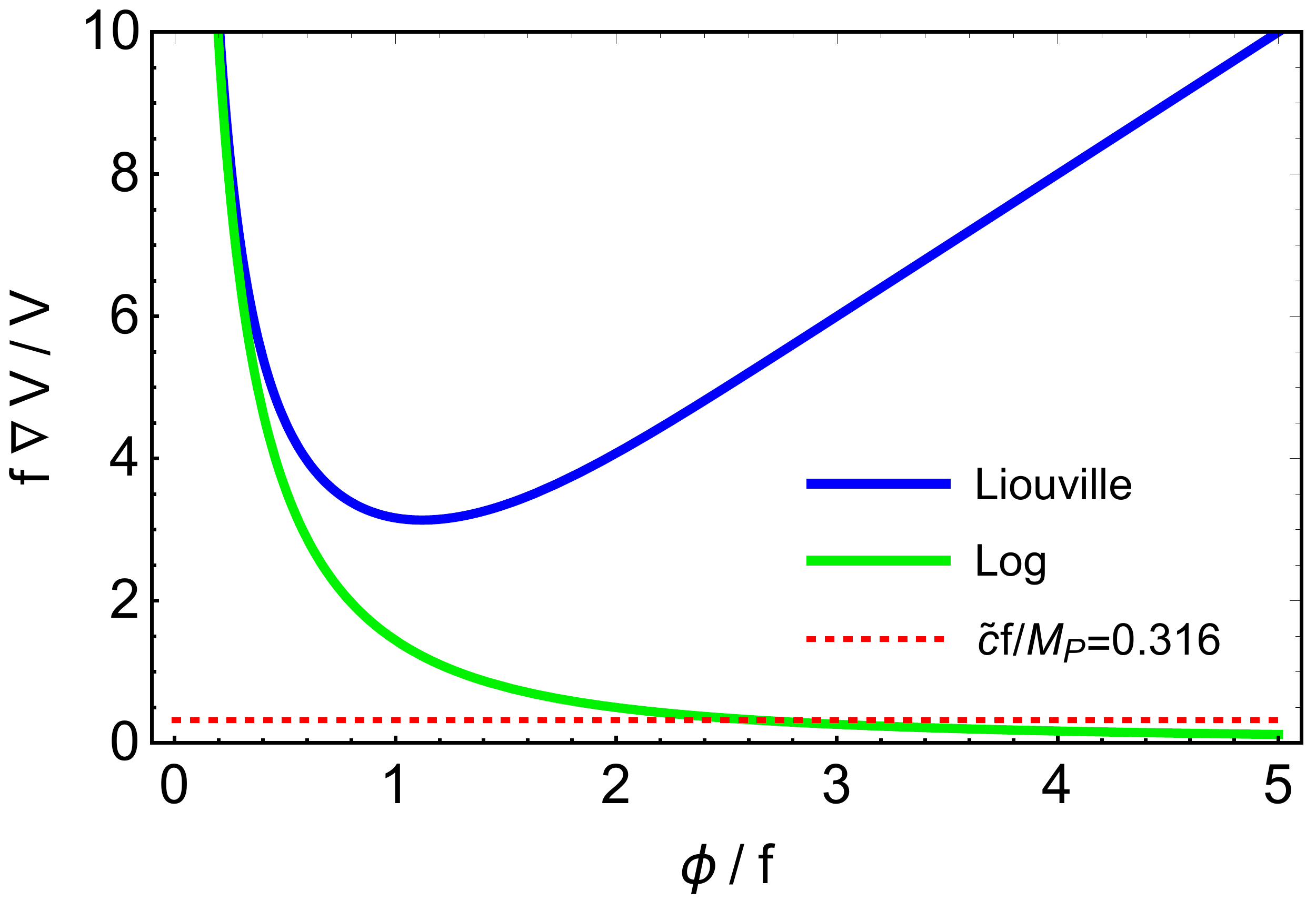}
\caption{Plot of $f|\partial_{|\Phi|}^{}\!V(|\Phi|)|/V$ as a function of $\phi\!/\!f$.\,
The red dashed line is $\tilde{ c}\!\sim\!\mathcal{O}(1)$ appearing in the Criterion\,2,
and $\tilde{c}\!=\!1$ is taken for our analysis. The potential $U_{{\rm Liouville}}^{}$ satisfies
the Criterion\,2 ($\MP|\partial_{|\Phi|}^{}\!V(|\Phi|)|/V\!\!>\!\tilde{c}$)
for all $\phi/f\!>\!0$,\, while $U_{{\rm Log}}^{}$ does so for $0\leqq\phi/\!f\leqq 2.6555$.}
\label{Fig:swampland}
\label{fig:6}
\end{figure}

To address the question for our study, we apply the following two criteria
\cite{Agrawal:2018own}:
\begin{itemize}
\item Criterion\,1: ~$\Delta\Phi \lesssim\Delta$\,,
\item Criterion\,2: ~$\MP |\nabla_{\Phi}^{}V|/V\geqq \tilde{c}$\,,
\end{itemize}

\hspace*{-9.9mm}
where $\Delta\Phi$ in Criterion\,1 is the difference between the maximum and the minimum
of the field value in unit of $\MP$.\,
Although the exact values of ($\Delta,\tilde{c}$) have not been determined yet, both are expected to be
$\mathcal{O}(1)$. Following~\cite{Herdeiro:2018hfp}, we take $\Delta$ and $\tilde{c}$ to be the unity for our analysis.

\vspace*{1mm}

For the complex scalar field in our case, the Criterion\,2
reads $\,\MP |\partial_{|\Phi|}^{}\!V(|\Phi|)|/V\!\geqq \tilde{c}$\,.\,
Using the two potential forms (\ref{VLiouville}) and (\ref{VLog}),
we can recast this condition as
\beqa
F(x) = \frac{2xe^{x^{2}}}{\,e^{x^{2}}\!-\!1\,}\geqq \frac{\tilde{c}\,f}{\,\MP\,}\,,
\quad\quad
G(x) = \frac{2x}{\,(1\!+\!x^{2})\log(x^{2}\!+\!1)}\geqq \frac{\tilde{c}\,f}{\,\MP\,}\,,
\label{eq:criterion2}
\eeqa
%
%\hspace*{-2.5mm}
where $\,x=\phi/f$,\, and $F(x)$ and $G(x)$ equal
$f|\partial_{|\Phi|}^{}\!V(|\Phi|)|/V$\, for $U_{\text{Liouville}}^{}$
and $U_{\text{Log}}^{}$ potentials, respectively.
The largest coupling strength parameter with which we apply this test is
$\,f=(1/\!\sqrt{10})\MP \!\simeq 0.316\MP$.\, So it suffices to check the inequalities
(\ref{eq:criterion2}) only for $\,f=(1/\!\sqrt{10})\MP$\,
as long as the Criterion\,2 is concerned.\footnote{One may apply a higher value
of the coupling strength parameter $\,f\,$ which corresponds to smaller self-interaction coupling regime $\Lambda<10$
in the right panel of  Fig.\,\ref{Compactness}.
Since we mainly want to check $C_{\max}^{}$ produced from strong coupling regime, we restrict our analysis to $f\leqq (1/\!\sqrt{10})M_{\rm P}$,\,
which corresponds to $\,\Lambda=1/\tilde{f}^{2}\geqq 10$.}
For the scalar field configuration corresponding to $C_{\max}^{}$,
if $F(x)$ and $G(x)$ conform to the inequalities
(\ref{eq:criterion2}) with $F(x),G(x)\!\geqq 0.316$,\,
then the Criterion\,2 is satisfied.
For the boson star study, we focus on the regular field configuration without a node
which continues to monotonically decrease from the center of the boson star and approaches
zero as moving outward. This implies $\Delta\Phi=\Phi(r\!=\!0)$.
For a fixed coupling strength parameter $f$,\,
we want to check whether the field configuration corresponding to
$C_{\max}^{}$ (as found in Sec.\,\ref{sec:numerical}) satisfies the
above Criteria\,1 and 2. If a violation occurs, then $C_{\max}^{}$
found in Sec.\,\ref{sec:numerical} is considered originated from an effective scalar theory belonging to the Swampland.

\vspace*{1mm}

In Fig.\,\ref{Fig:swampland}, we plot  $\,f|\partial_{|\Phi|}^{}\!V(|\Phi|)|/V$\,
as a function of $\,\phi/\!f$\,.\, We see that the Criterion\,2 is readily
satisfied by the Liouville potential for any $\phi/\!f\geqq 0$. This means that we only need to check whether $\phi(0)\!\leqq\! 1$ for $\phi(0)$ corresponding
to $M_{\max}^{}$ ($C_{\max}^{}$) with a given $f$.\,
As for the logarithmic potential, we note that $\,f|\partial_{|\Phi|}^{}\!V(|\Phi|)|/V$ is a monotonically decreasing function of $\phi/\!f$ and the Criterion\,2 is satisfied for
$\,0\leqq\phi/\!f\leqq 2.6555$.\, Hence, to avoid being in the Swampland,
we require the intersection between $\,\phi(0)\leqq 1$\, and $\,\phi(0)\leqq 2.6555f$\,
for a given $\,f\,$ to be satisfied by $\phi(0)$ value corresponding to $M_{{\max}}^{}$ ($C_{\max}^{}$).
We show the values of $\phi(0)/\!f$ associated with $C_{{\max}}^{}$ for each fixed $f$ in Table\,\ref{table:phif}, where the numbers in each parentheses correspond to the values of $\phi(0)$. In this table, we show the values of $\phi(r\!=\!0)$ up to three decimal points, which are enough for our purpose of checking two Swampland criteria.
For the Liouville potential, the inequality $\phi(0)<1$ holds for all $f$ values under consideration. For the logarithmic potential, both Criteria\,1 and 2 turn out to hold  for all $f$ values considered.
Hence, $C_{\max}^{{\rm Liouville}}$ and $C_{\max}^{{\rm Log}}$ in Fig.\,\ref{Compactness}
are all from the scalar theories in Landscape for $\Lambda\gtrsim 10$.

\tabcolsep 1pt
\begin{table}[t]
\centering\small
\begin{tabular}{c||c|c|c|c|c|c}
\hline\hline
&&&&&&
\\[-3mm]
$V(\Phi)$ & $f\!=\!1/\!\sqrt{10}$ & $f\!=\!1/\!\sqrt{20}$ & $f\!=\!1/\!\sqrt{40}$
& $f\!=\!1/\!\sqrt{60}$ & $f\!=\!1/\!\sqrt{80}$ & $f\!=\!1/\!\sqrt{100}$
\\
\hline\hline
&&&&&&
\\[-3mm]
Liouville & 1.866\,(0.59)    & 2.996\,(0.67) & 4.427\,(0.70)
& 5.593\,(0.722) & 6.494\,(0.726) & 7.3\,(0.73)
\\
\hline
&&&&&&
\\[-3mm]
~Logarithmic~ & ~0.727\,(0.23)~ & ~0.827\,(0.185)~ & ~0.885\,(0.14)~
& ~0.891\,(0.115)~ & ~0.939\,(0.105)~ & ~0.900\,(0.09)~
\\[0.5mm]
\hline\hline
\end{tabular}
\caption{The values of $\phi(r\!=\!0)/\!f$ corresponding to $M_{{\max}}^{}$ ($C_{{\max}}^{}$)
are presented for each potential and coupling strength parameter $f$ (in the unit of $\MP$).
In each parentheses the value of $\phi(0)$ is also shown. For the Liouville potential,
all the $\phi(0)$ values are less than 1 which is the intersection of the two criteria (see the text).
For the logarithmic potential, $\phi(r\!=\!0)/\!f<2.6555$ can be checked for $f$ considered
in this study, and so does $\phi(0)\!<\!1$.\, This shows that all the $C_{{\max}}^{}$ values shown in Fig.\,\ref{Compactness} can be regarded as arising from effective scalar theories UV-completed by a consistent quantum gravity.}
\label{table:phif}
\label{tab:1}
\end{table}

\vspace*{2mm}
\subsection{Semi-analytic Approach}
\label{semianalytic}
\vspace*{2mm}

In this subsection, we try to analytically understand the results we obtained in Sec.\,\ref{sec:numerical} by following the logic of \cite{Schiappacasse:2017ham,Croon:2018ybs,Guth:2014hsa,Chavanis:2017loo,Chavanis:2011zi}.
The strategy is to estimate the Hamiltonian of the system from Eq.(\ref{Ttt})
by using an ansatz for the radial profile of the field in Eq.(\ref{ansatz}).
This ansatz controls the shape of the profile by a single length scale $R$\,.\,
Following a variational approach, we extremize the Hamiltonian
with respect to $R$ and obtain an estimate of the maximum total mass $M_{{\max}}^{}$,\,
the minimum radius $R_{{\min}}^{}$, and the maximum compactness $C_{{\max}}^{}$.

\vspace*{1mm}

We first introduce an exponential ansatz of the scalar field wavefunction
by specifying $\Phi({\bold{r}},t)$ in Eq.(\ref{ansatz}) as
\beqa
\phi(r)=\sqrt{\!\frac{N}{\,\pi mR^{3}\,}\,}e^{-r/R},
\quad\quad
\omega^{2} \!=m^{2}\!\(\!1\!-\alpha\frac{\,\GN mN\,}{R}\!\) \!,
\label{ansatz2}
\eeqa
where we expect that the kinetic energy is proportional to the gravitational potential energy
up to a factor, which may be parametrized as $\,\alpha\,$ and will be determined later.
In the weak gravity limit, we may approximate the metric (\ref{metric}) as
\beqa
e^{\gamma(r)}\simeq 1+2V(r)\,,
\quad\quad
e^{\lambda(r)}\simeq 1-2V(r)\,,
\label{metric2}
\eeqa
where $\,V(r)\!=-G_{N}M(r)/r$\, is a gravitational potential.
Approximating the mass of the boson star system as $m_{\Phi}^{}N$,
we may use the following definition for the boson star mass,
\beqa
M^{A}(r)\equiv \int_{0}^{r}\!\!\dd r\,4\pi r^{2}\!
\(m_{\Phi}^{2}\Phi(r)^{2}\).
\label{totalMass}
\eeqa
Using this together with Eq.(\ref{ansatz2}), we obtain the gravitational potential
\beqa
V(r)=-\frac{\,\GN m_{\Phi}^{}N\,}{r}\!\left[1\!-\!e^{-\fr{2r}{R}}\!\!
\left(\!1\!+\!\frac{\,2r\,}{R}\!+\!\frac{\,2r^2\,}{R^{2}}\!\right)\!\right]\!.
\label{Gpotential}
\eeqa
\hspace*{-2.5mm}
The superscript $A$ in Eq.(\ref{totalMass}) stands for the mass to be used
in our analytic approach. We stress that this mass differs from the total mass
\eqref{massformula} that we used for the numerical computations in Sec.\ref{sec:numerical}.
From Eqs.(\ref{Ttt}) and (\ref{metric2}), we derive in the weak gravity limit,
\beqa
T^{0}_{0}(r) = -\omega^{2}[1\!-\!2V(r)]\phi^{2}-[1\!+\!2V(r)]
(\partial_{r}^{}\phi)^{2}-U(\phi^{2})\,,
\label{T00}
\eeqa
%
%\hspace*{-4mm}
which leads to the following Hamiltonian of the boson star when integrating over the proper volume\footnote{The integration measure should be $\dd^{3}r\sqrt{-g}$. For simplicity of the analysis,
we take the weak gravity limit and approximate $\sqrt{-g}\simeq 1$.}
\beqa
H_{{\rm tot}}^{} = \int \!\!\dd^{3}r\, T^{0}_{0}(r)\,.
\eeqa
We may decompose $H_{{\rm tot}}^{}$ into three different contributions,
\beqs
\beqa
H_{{\rm grav}}^{}\!+H_{{\rm mass}}^{}
&\,=\,&  -\!\int_{0}^{\infty}\!\!\dd^{3}r
\left\{\omega^{2}[1\!-\!2V(r)]\phi^2\!+ m_{\Phi}^{2}\phi^{2}\right\}\!,
\label{Hgrav}
\\[1mm]
\label{Hkin}
H_{{\rm kin}} &\,=\,&
-\!\int_{0}^{\infty}\!\!\dd^{3}r\, (\partial_{r}^{}\phi)^2 [1\!+\! 2V(r)] \,,
\\[1mm]
\label{Hint}
H_{{\rm int}} &\,=\,&
-\!\int_{0}^{\infty}\!\!\dd^{3}r\left[U(\phi^{2})\!-m_{\Phi}^{2}\phi^{2}\right]\!.
\eeqa
\eeqs
%
%\hspace*{-2mm}
By substituting the explicit form of the ansatz (\ref{ansatz2}) into Eq.(\ref{Hgrav}), we obtain
\beqa
H_{{\rm grav}}
+H_{{\rm mass}}=-2m_{\Phi}N+\frac{\,\GN m_{\Phi}^{2}N^{2}(8\alpha\!-\!5)\,}{8R}\,,
\label{Hgrav2}
\eeqa
where we have dropped an additional term $\,5G_{N}^{2}m_{\Phi}^{3}N^{3}\alpha/8R^{2}$\,
due to its large suppression by the squared Newton constant $G_{N}^2$.\,
We expect Eq.(\ref{Hgrav2}) to reduce to $H_{{\rm grav}}^{}$
in the non-relativistic limit and this requirement fixes the value
$\alpha =\fr{5}{4}$.\footnote{We refer the non-relativistic-limit result
$H_{\rm grav}^{}=-5\GN m_{\Phi}^{2}N^{2}/16R$ to Ref.\,\cite{Schiappacasse:2017ham}.}
Thus, we obtain
\beqa
H_{{\rm grav}} + H_{{\rm mass}}
= -2m_{\Phi}N+\frac{5\GN m_{\Phi}^{2}N^{2}}{8R}\,.
\label{Hgrav3}
\eeqa

Similarly, we derive the kinetic energy arising from the gradient of the field,
\beqa
H_{{\rm kin}}^{}
= -\frac{N}{\,m_{\Phi}^{}R^{2}\,} +
\frac{\,5\GN N^{2}\,}{8R^{3}}\,.
\eeqa
Finally, we compute $H_{{\rm int}}^{}$ due to the two different potentials
introduced in Sec.\,\ref{sec:potential}.
For the Liouville potential (\ref{VLiouville}), the exponential ansatz leads to
\begin{eqnarray}
H_{{\rm int}}^{{\rm Liouville}} &=&
-\int_{0}^{\infty}\!\!\dd r\,4\pi r^{2}
\left[U(\phi^{2})-m_{\Phi}^{2}\phi^{2}\right]
\nn\\
&=&
-f^{2}m_{\Phi}^{2}\int_{0}^{\infty}\!\!\dd r\,4\pi r^{2}
\sum_{k=2}^{\infty}\frac{\phi^{2k}}{\,f^{2k}k!\,}
\nn\\
&=&
-4\pi f^{2}m_{\Phi}^{2}\sum_{k=2}^{\infty}\frac{1}{\,f^{2k}k!\,}
\(\!\frac{N}{\,\pi m_{\Phi}^{}R^{3}\,}\!\!\)^{\!\!\!k}\!\!
\int_{0}^{\infty}\!\! \dd r\,r^{2}e^{-\fr{2kr}{R}}
\nn\\
&=&
-\pi f^{2}m_{\Phi}^{2}R^{3}\sum_{k=2}^{\infty}\frac{1}{\,f^{2k}k!k^{3}\,}
\(\!\!\frac{N}{\,\pi m_{\Phi}^{}R^{3}\,}\!\!\)^{\!\!\!k} ,
\label{HLiouville}
\end{eqnarray}
which provides a series of repulsive forces.
For the logarithmic potential (\ref{VLog}), the exponential ansatz leads to
\begin{eqnarray}
H_{{\rm int}}^{{\rm Log}}&=&
-\int_{0}^{\infty}\!\!\dd r\,4\pi r^{2}
\left[U(\phi^{2})-m_{\Phi}^{2}\phi^{2}\right]
\nn\\
&=& -f^{2}m_{\Phi}^{2}\int_{0}^{\infty}\!\!\dd r\,4\pi r^{2}\!
\sum_{k=2}^{\infty}(-1)^{k-1}\frac{\phi^{2k}}{kf^{2k}}
\nn\\
&=& -4\pi f^{2}m_{\Phi}^{2}\sum_{k=2}^{\infty}\frac{(-1)^{k-1}}{kf^{2k}}\!
\(\!\!\frac{N}{\,\pi m_{\Phi}R^{3}\,}\!\!\)^{\!\!\!k}\!\!
\int_{0}^{\infty}\!\!\dd r\,r^{2}e^{-\fr{2kr}{R}}
\nn\\
&=&
-\pi f^{2}m_{\Phi}^{2}R^{3}\sum_{k=2}^{\infty}\frac{(-1)^{k-1}}{k^{4}f^{2k}}\!
\(\!\!\frac{N}{\,\pi m_{\Phi}R^{3}\,}\!\!\)^{\!\!\!k}\,,
\end{eqnarray}
which yields a series of pairs of attractive and repulsive forces.

\vspace*{1mm}

Given the above explicit expressions for the different self-interaction Hamiltonians, we can estimate minimum size of the scale radius $R_{{\rm min}}^{}$ and an associated $N_{{\max}}^{}$ by extremizing $H_{{\rm tot}}^{}$.
As shown in \cite{Schiappacasse:2017ham}, it is convenient to go through the extremizing procedure after rescaling $R$, $N$ and $H$ into the dimensionless quantities,
\beqa
\tilde{R}\equiv m_{\Phi}f\sqrt{\GN}R\,, \quad\quad
\tilde{N}\equiv \frac{m_{\Phi}^{2}\!\sqrt{\GN}}{f}N\,,
\quad\quad
\tilde{H}\equiv\frac{m_{\Phi}}{\,f^{3}\!\sqrt{\GN}\,}H\,.
\label{rescaling}
\eeqa
This results in
{\small
\beqs
\beqa
\tilde{H}^{}_{{\rm Liouville}} &=&
-\frac{2\tilde{N}}{f^{2}G_{N}}+\frac{5\tilde{N}^{2}}{8\tilde{R}}
-\frac{\tilde{N}}{\tilde{R}^{2}}+\frac{5(f^{2}G_{N})\tilde{N}^{2}}{8\tilde{R}^{3}}
-\sum_{k=2}^{\infty}\frac{(f^{2}G_{N})^{k-2}}{k^{3}k!\pi^{k-1}}\frac{\tilde{N}^{k}}{\tilde{R}^{3k-3}}\,,
\hspace*{10mm}
%\eeqa
%and
%\beqa
%
\\[1mm]
\tilde{H}^{}_{{\rm Log}} &=&
-\frac{2\tilde{N}}{f^{2}G_{N}}+\frac{5\tilde{N}^{2}}{8\tilde{R}}
-\frac{\tilde{N}}{\tilde{R}^{2}}+\frac{5(f^{2}G_{N})\tilde{N}^{2}}{8\tilde{R}^{3}}
-\sum_{k=2}^{\infty}\frac{(-1)^{k-1}(f^{2}G_{N})^{k-2}}{k^{4}\pi^{k-1}}
\frac{\tilde{N}^{k}}{\tilde{R}^{3k-3}}\,.
\hspace*{16mm}
\eeqa
\eeqs
}

\begin{figure}[t]
\centering
\includegraphics[scale=0.45]{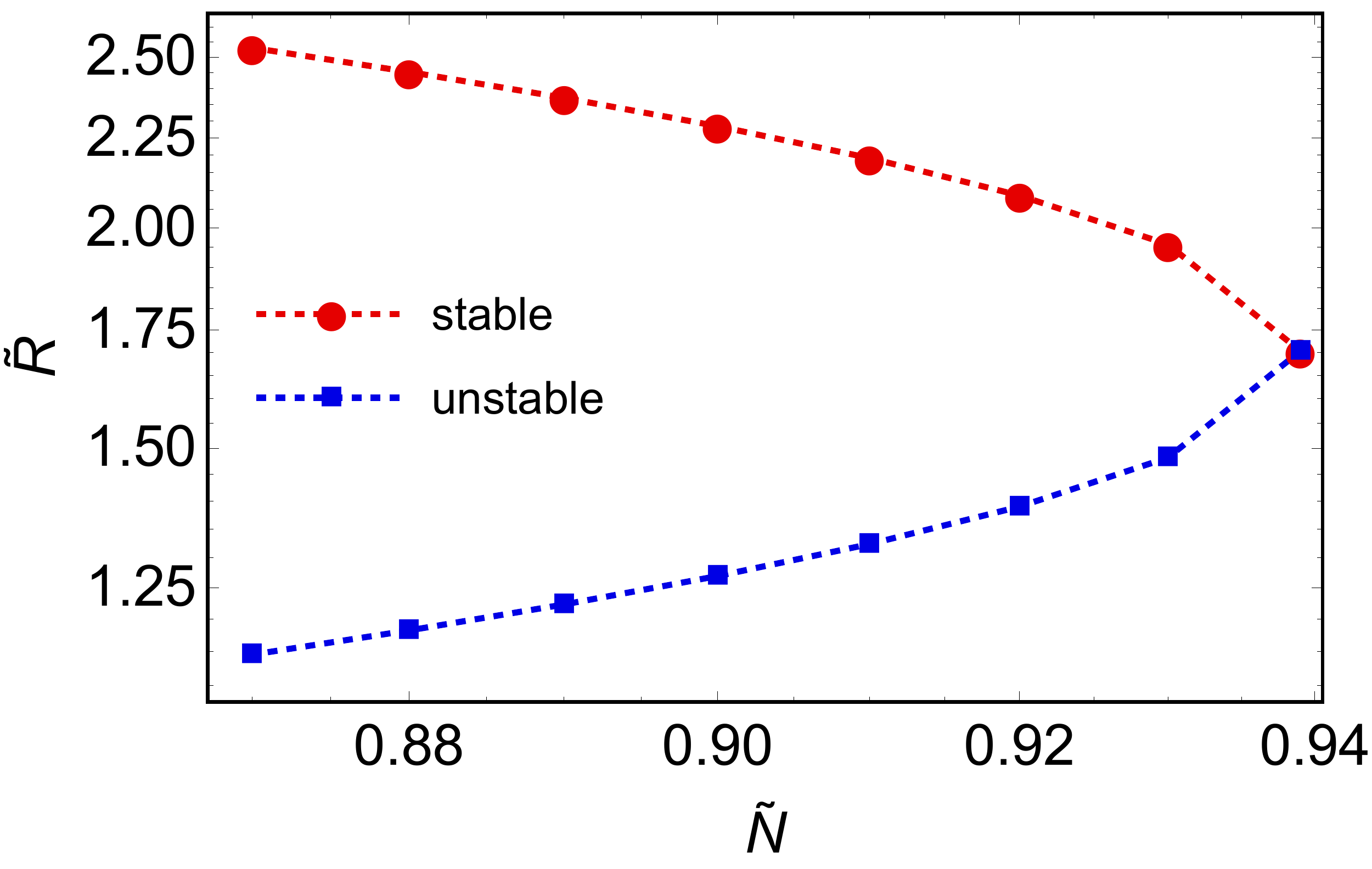}
\vspace*{-2mm}
\caption{Solutions of $\,\partial\tilde{H}\!/\partial\tilde{R}\!=\!0\,$
for $\,f^{2}\GN =1$\, and for the Liouville potential with truncation at $\phi^{6}$ term.
For a fixed value of $\tilde{N}$, there exists a pair of $\tilde{R}$\,.\,
The larger $\tilde{R}$ makes $\partial^{2}\!\tilde{H}\!/\partial\!\tilde{R}^{2}>0$\,
(red curve, stable branch) and the smaller one makes
$\partial^{2}\tilde{H}/\partial\tilde{R}^{2}<0$
(blue curve, unstable branch).
For the $\tilde{N}$ value at which the two different branches meet together,
the boson star realizes its highest possible compactness.
For $(\tilde{N}_{\max}^{},\tilde{R}_{\min}^{})
=(0.939007500937,1.70293)$, the relevant boson star is characterized by $\,C_{\max}^{\rm A}\!=0.129882$.
}
\label{Fig:analytic_NvsR}
\label{fig:7}
%\vspace*{0.5mm}
\end{figure}

Next, for a set of values of
$f^{2}\GN\simeq(f/m_{{\rm P}}^{})^{2}$,\,
we estimate a minimum radius $\tilde{R}_{{\min}}^{}$
and a maximum number of constituent bosonic particles
$\tilde{N}_{{\max}}^{}$ which can be realized for a stable boson star.
For this purpose, we take a fixed value of $\tilde{N}$ and we search for $\tilde{R}$ associated with a local minimum and maximum of $\tilde{H}$, which correspond to the stable and unstable boson star branches, respectively.

\vspace*{1mm}

\begin{figure}[t]
\centering
\includegraphics[scale=0.45]{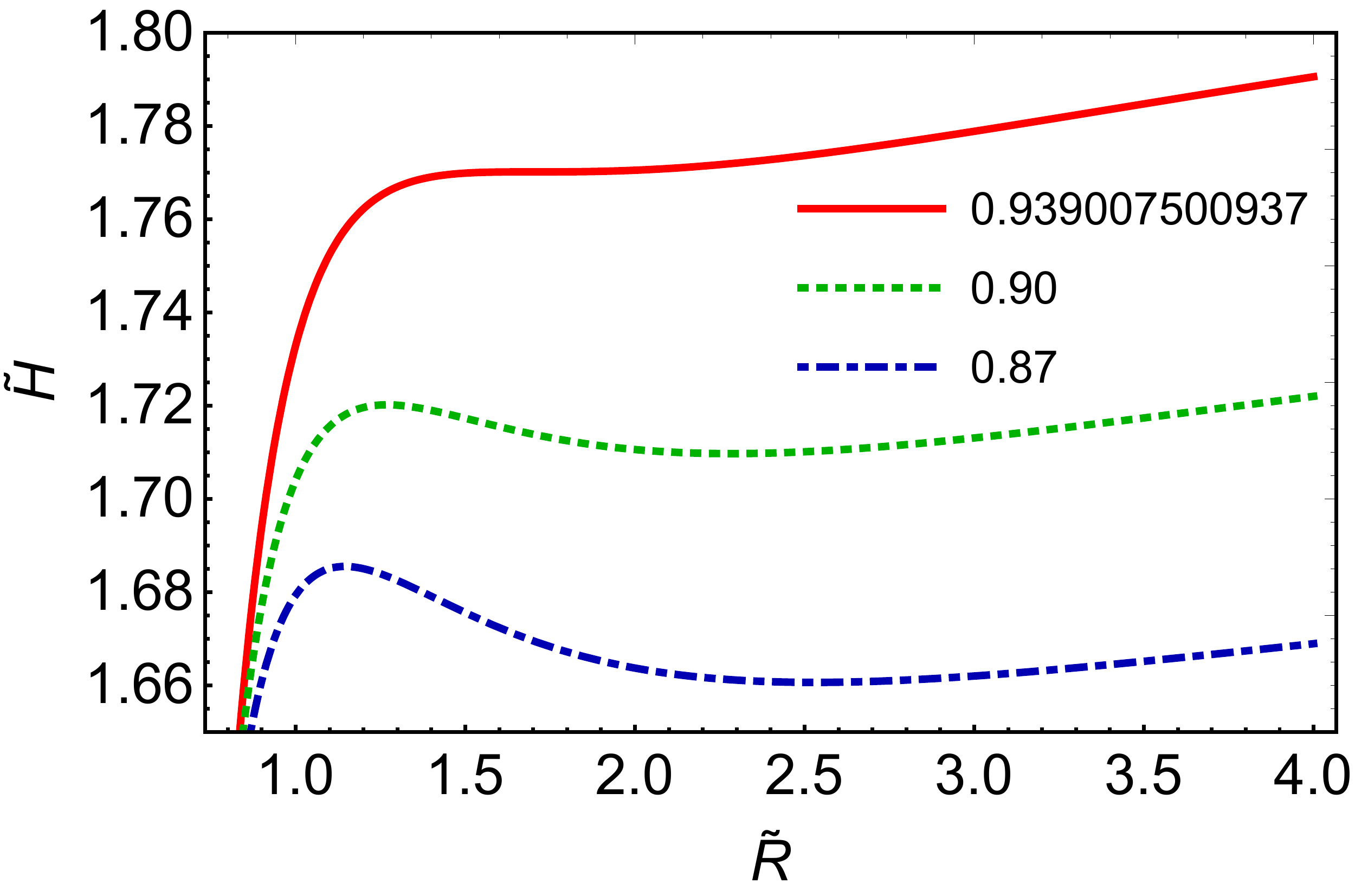}
\vspace*{-2mm}
\caption{Plot of $\,\tilde{H}\,$ as a function of $\,\tilde{R}\,$ for $\,f^{2}\GN\!=\!1$\,
and for the Liouville potential truncated at $\phi^{6}$ term. The red solid curve
is for $\,\tilde{N}\!=\!0.939007500937$\,,\, the green dashed curve for
$\,\tilde{N}=0.90$,\, and the blue dash-dotted curve for $\,\tilde{N}=0.87$.\,
In each case, there exist two extremum and those get closer to each other
as $\tilde{N}$ increases. Eventually for $\tilde{N}\!=0.939007500937$, those two merge
into a single one which is identified as $\tilde{R}_{\min}^{}$.}
\label{Fig:analytic_HvsR}
\label{fig:8}
\end{figure}

In Fig.\,\ref{Fig:analytic_NvsR}, we present an example of how to
find $(\tilde{R}_{\min}^{},\tilde{N}_{\max}^{})$ from Taylor-expanded Liouville potential with a truncation at $\phi^{6}$ term and $f^{2}\GN \!=\!1$\,.\,
For a fixed value of $\tilde{N}$, we find a pair of $\tilde{R}$'s satisfying $\partial\tilde{H}\!/\!\partial\tilde{R}=0$\,.\,
The $\tilde{R}$ solution obeying $\,\partial^{2}\!\tilde{H}/\partial\!\tilde{R}^{2}\!>\!0$\,
corresponds to the stable boson star solution and vice versa.
For the high enough $\tilde{N}$, two branches merge into a single solution of
$(\tilde{N},\tilde{R})$ eventually, which yields
$\tilde{N}_{{\max}}^{}$ and $\tilde{R}_{{\min}}^{}$.\,
In Fig.\,\ref{Fig:analytic_HvsR}, we illustrate how the two extremum
of $\tilde{H}$ approach each other as $\tilde{N}$ increases.

\vspace*{1mm}

Given $(\tilde{N}_{{\max}}^{},\tilde{R}_{{\min}}^{})$ for a choice of $f^{2}\GN$ and using Eq.(\ref{Cdef}),
 we compute the maximum compactness which the boson star can realize,
\beqa
C_{{\max}}^{\rm A}
=\frac{\,\GN 0.99 M^{\rm A}\,}{R_{99}^{}}
=\frac{\,\GN\,0.99 m_{\Phi}N_{{\max}}^{}\,}{4.203R_{{\min}}^{}}
=\frac{0.99}{4.203}\!\times\!\frac{\,f^{2}\GN\tilde{N}_{\max}^{}\,}{\tilde{R}_{\min}^{}}\,,
\hspace*{12mm}
\eeqa
where the superscript on $C_{\max}^{}$ means that it is computed with
$M^{\rm A}$ defined in Eq.(\ref{totalMass}) and the factor of 4.203 arises from
\beqa
0.99N =\int_{0}^{4.203R}\!\!\dd r\,4\pi r^{2} \!\left[m_{\Phi}^{}\Phi^{2}(r)\right] .
\eeqa
For instance, given
$(\tilde{N}_{\max}^{},\,\tilde{R}_{\min}^{})
 =(0.939007500937,\,1.70293)$
as in Fig.\,\ref{Fig:analytic_NvsR}, we deduce $C_{\max}^{}=0.129882$\,.

\vspace*{1mm}

Note that for an estimate of compactness by the semi-analytic approach, we approximate the boson star mass as $m_{\Phi}N$ for simplicity, though this does not take into account the binding energy of the boson star. In the following, we will compare $C_{\rm max}^{A}$ to $C_{{\rm max}}$ from the complete numerical computation in order to understand different binding energy behaviors of boson stars of different potential observed in Sec.~\ref{sec:numerical} for the small coupling regime.

\vspace*{1mm}

With the semi-analytic approach described above, we try to understand the numerical results obtained in Sec.\,\ref{sec:numerical}.
For the maximum compactness in Fig.\,\ref{Compactness},
we observe that starting from the same value in free theory limit
($1\!/\!\tilde{f}^{2}\!=\!0$),
$C_{\max}^{}$ due to $U_{{\rm Liouville}}^{}$ and $U_{{\rm Log}}$
splits up with the increasing self-interaction strength and ends up with the hierarchy,
\beqa
C_{{\max}}^{{\rm Liouville}}>C_{\max}^{{\rm Quartic}} > C_{\max}^{{\rm Log}} \,,
\eeqa
where $C_{\max}^{{\rm Quartic}}$ is due to the repulsive quartic self-interaction ($\sim\!\!\lambda_\Phi^{}\Phi^{4}$ with $\,\lambda_\Phi^{}\!\!>\!0$).
This shows that an infinite number of repulsive self-interactions from expansion of
$U_{{\rm Liouville}}^{}(|\Phi|^{2})$
renders the boson star system more compact than that with the usual repulsive
quartic self-interaction. On the other hand, alternating attractive and repulsive
self-interactions arising from expanding $U_{{\rm Log}}^{}(|\Phi|^{2})$
tend to cancel each other and yield a small net attractive non-gravitational force. Hence, this small net effect makes the relevant boson star less compact than that with
the repulsive self-interaction.

\vspace*{1mm}

\begin{figure}[t]
\centering
\hspace*{-5mm}
\includegraphics[width=7.7cm,height=5.9cm]{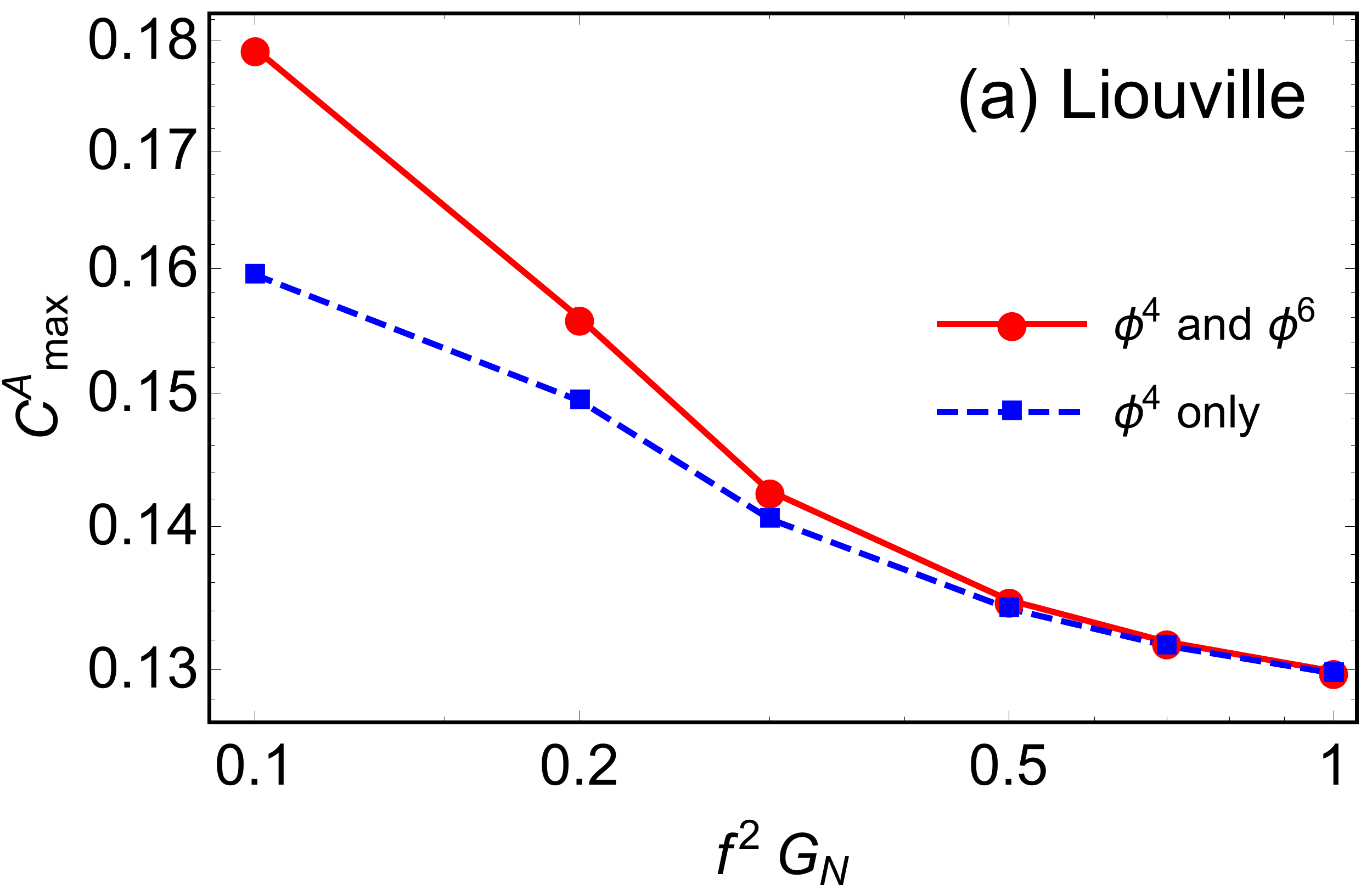}
\includegraphics[width=7.7cm,height=5.9cm]{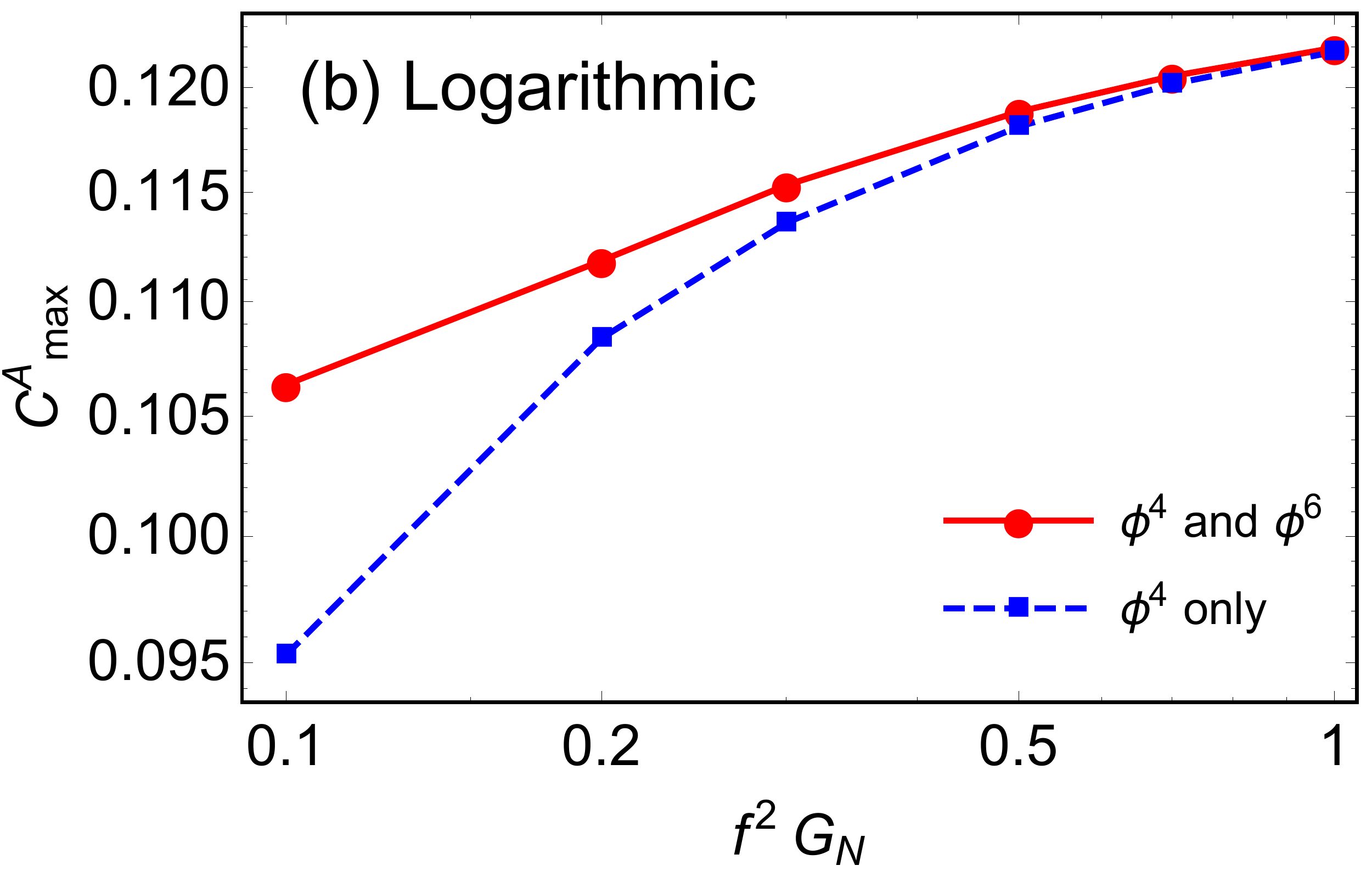}
\caption{Analytic computation of $C_{{\max}}^{\rm A}$
from the Taylor-expanded Liouville potential and logarithmic potential with truncation at $\Phi^{4}$ and $\Phi^{6}$ terms.}
\vspace*{-1.5mm}
\label{fig:analytic_Cmax}
\label{fig:9}
\end{figure}

In Fig.\,\ref{fig:analytic_Cmax}, we plot $C_{\max}^{\rm A}$ as a function of $f^2\GN$,\, where $C_{{\rm max}}^{\rm A}$ is
defined in Eq.(\ref{Cdef}) following the semi-analytic approach explained above.
The left and right panels show $C_{{\max}}^{\rm A}$ based on the
Taylor-expanded Liouville and logarithmic potentials truncated at $\Phi^{4}$ and $\Phi^{6}$
terms, respectively. For Liouville potential, both $\Phi^{4}$ and $\Phi^{6}$ terms
have positive signs and thus generate repulsive forces. The $\Phi^{6}$ contribution to the repulsive force together with that of the $\Phi^{4}$ term
helps for competing with the gravitational attraction, which leads to the higher compactness
for a fixed $f^2\GN$.\,
This tendency becomes more prominent as self-interaction strength increases
(as $f^2\GN$ decreases). When there are more additional repulsive forces
provided by the rest of the higher order terms with positive sign
in the Taylor-expanded Liouville potential,
we expect the boson star would become more compact as compared
to the case of the usual repulsive $\Phi^{4}$ only.
This is well reflected in Fig.\,\ref{Compactness} with the blue dashed curve
($C_{{\max}}^{}$ due to complete Liouville potential)
and green dashed curve ($C_{{\max}}^{}$ due to repulsive quartic interaction).
In contrast, for the logarithmic potential, $\Phi^{4}$ term produces attractive
force whereas $\Phi^{6}$ term is repulsive.
Accordingly, the repulsive $\Phi^{6}$ term plays a role of
diluting the attractive force of the $\Phi^{4}$ term.
The attractive $\Phi^{4}$ term tends to decrease $C_{\max}^{\rm A}$
as compared to a boson star with a quartic repulsive self-interaction
in a free scalar theory, and this becomes more significant for the
higher self-interaction strength (for the smaller $f^2\GN$). But, the degree of decrease in $C_{{\max}}^{\rm A}$ becomes weaker when the repulsive force of $\Phi^{6}$ term dilutes the attractive force from $\Phi^{4}$. In the end, we expect the net attractive force
due to pairing attractive and repulsive forces from terms in Taylor-expanded logarithmic
potential to make $C_{\max}^{{\rm A}}$ lie between 0.095 and 0.105
at $f^{2}\GN =0.1$.\footnote{We have checked that $C_{\max}^{\rm A}$ from interaction terms up to $\Phi^{10}$ in expansion of logarithmic potential
takes a value of $\,C_{\max}^{\rm A}\!=0.103$\,.}

\vspace*{1mm}

According to the definition of $C_{\max}^{\rm A}$,
we may infer the behavior of the negative binding energy
$E_B^{}$ from $C_{\max}^{\rm N}\!-C_{\max}^{\rm A}\!\!\propto\! E_B^{}$\,,\,
where $C_{\max}^{\rm N}$ is the maximum compactness computed numerically
in Sec.\,\ref{sec:numerical}.
The value $\,f^{2}\GN\!=\!0.1$\, corresponds to the coupling strength
$\,\Lambda\simeq\!0.4\,$,\,  so changing from $\,f^{2}\GN\!=\!1$\,
to $\,f^{2}\GN\!=\!0.1$\, corresponds to varying the coupling strength from
$\,\Lambda\simeq\!0.04\,$ to $\,\Lambda\simeq\!0.4\,$.\,
For Liouville potential, we observe that as the coupling $\,\Lambda$ rises, both
$\,C_{\max}^{\rm N}$ and $C_{\max}^{\rm A}$\, increase, and
the difference $|C_{\max}^{\rm N}\!\!-\!C_{\rm max}^{\rm A}|\!\!\propto\!\! |E_B^{}|$
also rises in the small coupling regime.
This trend is in agreement with the red curve in the left panel of
Fig.\,\ref{Compactness}. On the other hand, for the logarithmic potential,
$C_{\max}^{A}$ decreases and $C_{\max}^{N}$ increases for rising $\,\Lambda$
in the small coupling regime where $C_{\max}^{A}>C_{\max}^{N}$,\,
so the difference
$|C_{\max}^{\rm N}\!\!-\!C_{\rm max}^{\rm A}|\!\propto\! |E_B^{}|$
decreases for the small coupling regime.
This explains the behavior of the blue curve in the small coupling regime as shown
by the left panel of Fig.\,\ref{Compactness}.

\vspace*{2mm}
\section{Astrophysical Probe of Boson Stars}
\label{sec:AP}
\label{sec:4}

In this section, we study two ways to experimentally probe the presence of boson stars
by constraining their model parameter space. The boson stars may be
responsible for a small fraction of the dark matter in the Universe by serving as a kind of
MACHO (macroscopic compact halo object), which we infer from  \cite{Tisserand:2006zx,Griest:2013aaa,Niikura:2017zjd,Brandt:2016aco,Allsman:2000kg}.
For the present study, we focus on the scenarios
where the fraction of the DM provided by the boson stars is either less than $10\%$ or $1\%$.

\vspace*{1mm}

In Sec.\,\ref{sec:scalarmass}, we will derive the allowed mass range of the scalar particles
which compose the boson stars of our interest.
In Sec.\,\ref{sec:FRB}, we consider that a small fraction of the dark matter is attributed to
the boson stars, and study how we can probe a boson star by the lensing of fast radio burst (FRB).
Then, in Sec.\,\ref{sec:GW}, we analyze whether the gravitational wave (GW) signals caused
by the merger of two boson stars can be detected by the laser interferometer
gravitational-wave observatory (LIGO).
For both astrophysical probes, we will discuss the different implications for the
boson stars with the two benchmark scalar potentials as studied earlier.

\vspace*{2mm}
\subsection{Scalar Particle Mass Range for Boson Stars}
\label{sec:scalarmass}
\label{sec:4.1}
\vspace*{2mm}

In Sec.\,\ref{sec:numerical}, we computed the maximum total mass $M_{{\rm max}}^{}$
of a boson star in unit of $\,m_{{\rm P}}^2/m_{\Phi}^{}$,\, for each scalar potential with a coupling strength $\Lambda$. Given a fraction of the DM contributed by the boson stars ($\xi_{{\rm DM}}^{}$), we can derive the relevant mass range of the scalar particle $\Phi$
by comparing $M_{{\rm max}}^{}$ to the current constraint on the MACHO mass.

\vspace*{1mm}

The microlensing survey provides constraints on $\xi_{{\rm DM}}^{}$ and $M_{{\max}}^{}$
for MACHO mass range $\,10^{-11}\!<M/M_{\odot}^{}\!<30$\,  \cite{Tisserand:2006zx,Griest:2013aaa,Niikura:2017zjd,Allsman:2000kg},
which can be applied to the primordial black holes (PBHs) and the exotic compact objects including boson stars.
For the heavier mass range of $\,M\!\gtrsim\! 100M_{\odot}^{}\,$,\,
the cosmic microwave background (CMB) anisotropy excludes PBHs
as the dominant component of DM \cite{Ali-Haimoud:2016mbv}.
On the other hand, the survival of a star cluster near the core of Eridanus\,II
and of a sample of compact ultra-faint dwarfs places constraints on
$\xi_{{\rm DM}}^{}$ and $M_{{\rm max}}^{}$ for MACHO mass range $\,M\!\gtrsim 5M_{\odot}^{}$ \cite{Brandt:2016aco}.
For exemplary fractions $\,\xi_{{\rm DM}}^{}=0.1$ and $0.01$,\,
we find the allowed MACHO mass ranges to be
$\,1\lesssim M/M_{\odot}^{}\!\lesssim\!100$\, and
$\,10^{-7}\!\lesssim\! M/M_{\odot}^{}\!\lesssim\!10^{3}$,\,
respectively. In Table\,\ref{table:phimass}, for $\Lambda\!=\!10$ and $\Lambda\!=\!100$\,,\, we derive
the corresponding constraints on the scalar boson mass for the two benchmark potentials,
which is obtained based on the analysis of $M_{{\max}}^{}$ in
Sec.\,\ref{sec:numerical}.
\tabcolsep 1pt
\begin{table}[t]
\centering\small
\begin{tabular}{c||c|c}
\hline\hline
&&
\\[-3mm]
~~$\xi_{{\rm DM}}^{}$~~
& ~Liouville Potential\,($\Lambda\!=\!10$)~
& ~Logarithmic Potential\,($\Lambda\!=\!10$)~
\\
\hline\hline
&&
\\[-3mm]
$0.1$ & ~$1.3\!\times\!10^{-12}\lesssim m_{\Phi}^{}\!\lesssim 1.3\!\times\!10^{-10}~$
& $~6.4\!\times\!10^{-13}\lesssim m_{\Phi}^{}\!\lesssim 6.4\!\times\!10^{-11}$~
\\
\hline
&&
\\[-3mm]
$0.01$ & $1.3\!\times\!10^{-13}\lesssim m_{\Phi}^{}\!\lesssim 1.3\!\times\!10^{-3}$
& $6.4\!\times\!10^{-14}\lesssim m_{\Phi}^{}\!\lesssim 6.4\!\times\!10^{-4}$
\\[0.5mm]
\hline\hline
&&
\\[-3mm]
~~$\xi_{{\rm DM}}^{}$~~
& ~Liouville Potential\,($\Lambda\!=\!100$)~
& ~Logarithmic Potential\,($\Lambda\!=\!100$)~
\\
\hline\hline
&&
\\[-3mm]
$0.1$ & ~$3.6\!\times\!10^{-12}\lesssim m_{\Phi}^{}\!\lesssim 3.6\!\times\!10^{-10}~$
& $~8.4\!\times\!10^{-13}\lesssim m_{\Phi}^{}\!\lesssim 8.4\!\times\!10^{-11}$~
\\
\hline
&&
\\[-3mm]
$0.01$ & $3.6\!\times\!10^{-13}\lesssim m_{\Phi}^{}\!\lesssim 3.6\!\times\!10^{-3}$
& $8.4\!\times\!10^{-14}\lesssim m_{\Phi}^{}\!\lesssim 8.4\!\times\!10^{-4}$
\\
\hline\hline
\end{tabular}
\caption{Allowed mass ranges of the scalar particle under two benchmark potentials,
as inferred from the MACHO mass constraints\,\cite{Tisserand:2006zx,Griest:2013aaa,Niikura:2017zjd,Brandt:2016aco,Allsman:2000kg} for the DM fraction $\xi_{{\rm DM}}^{}\!=0.1,0.01$ and the coupling strength
$\Lambda = 10,100$.
Here the unit of the scalar particle mass is eV.}
\label{table:phimass}
\label{tab:2}
\end{table}

\vspace*{1mm}

In addition, we may wonder whether the scalar mass range as allowed by the MACHO constraints
(Table\,\ref{tab:2}) can be consistent with the cold dark matter (CDM) isocurvature modes
in the CMB power spectrum. For the present boson star study, we consider the scenario where the scalar particle mass is comparable to
the Hubble expansion rate during inflation ($H_{\rm inf}^{}$) \cite{Bertolami:2016ywc}. During inflation the scalar mass $m_{\Phi,{\rm inf}}^{}\simeq H_{{\rm inf}}^{}$
can be generated by the gravitationally induced coupling between
the scalar boson $\Phi$ and inflaton $\chi$ \cite{Bertolami:2016ywc},
\beqa
\mathcal{L}_{{\rm \Phi\chi}}
= c_{{\rm \Phi\chi}}^{}\frac{\,V(\chi)\,}{M_{{\rm P}}^{2}}\Phi^{2}\,,
\label{eq:scalarinflaton}
\eeqa
where $V(\chi)$\, is the inflaton potential
and $c_{{\rm \Phi\chi}}^{}$ is a dimensionless coupling parameter.
Since the scalar mass of our interest obeys $\,m_{\Phi}^{}\!\ll\! H_{{\rm inf}}^{}$\,,\,
the $\Phi$ mass during inflation ($m_{\Phi,{\rm inf}}^{}$) becomes effectively
$m_{\Phi,{\rm inf}}^{}\simeq H_{{\rm inf}}^{}$\,.\,
For the post-inflationary epoch, $m_{\Phi,{\rm inf}}^{}$ reduces to the original
small mass ($m_{\Phi}^{}$) which corresponds to $\,V(\chi)\!=0$ as the global minimum of the inflaton potential.

\vspace*{1mm}

For the two benchmark scalar potentials in Sec.\,\ref{sec:potential},
their mass term dominates over the quartic interaction and non-renormalizable terms arising from expanding the potential with $\,|\Phi|\!<f$\,.\, Hence,
after inflation ends, the homogeneous scalar field obeys the time evolution equation
in the expanding background,
\beqa
\ddot{\Phi}+3H\dot{\Phi}+m^{2}_{\Phi}\Phi \,=\, 0\,,
\label{eq:Tevolution}
\eeqa
where $H$ is the Hubble expansion rate and the dot denotes
the derivative with respect to time $t$\,.
For the scalar mass $\,m_{\Phi}^{}\!<\!\mathcal{O}(10^{-3})$eV\,
relevant to the boson stars serving as a candidate of MACHO,
the relation $\,H\!>\! m_{\Phi}^{}$\, holds after inflation
and the field value remains as $\phi_{{\rm inf}}^{}$
(the field displacement from the global minimum of the potential)
due to the overdamping from the end of the inflation to the time
when $\,H\simeq m_{\Phi}^{}$\, is reached.
Accordingly, the energy density of the scalar particles
remains as a constant, $\,\rho_{\Phi}^{}=m^{2}_{\Phi}\phi^{2}_{{\rm inf}}/2\,$,
until $\,H\simeq m_{\Phi}^{}\,$ is realized.
Then, $\rho_{\Phi}^{}$ scales as $a^{-3}$.
When the oscillation starts, we have $H\!\simeq m_{\Phi}^{}$
and the temperature of the Universe is given by
\beqa
T_{{\rm osc}}^{}=\(\!\frac{90}{\pi^{2}}\!\)^{\!\!\!1/4}\!\!
g_{*}^{1/4}\sqrt{M_{\rm P}^{}m_{\Phi}^{}\,}\,,
\label{eq:Tosc}
\eeqa
where $g_{*}^{}$ is the effective number of relativistic degrees of freedom.
In our scenario, we denote the fraction of DM contributed by the boson stars as $\xi_{{\rm DM}}^{}$.
With $T_{{\rm osc}}^{}$ in Eq.(\ref{eq:Tosc}) for $n_{\Phi}^{}/s$,
we derive the relation\,\cite{Bertolami:2016ywc},
\beqs
\beqa
\xi_{{\rm DM}}^{}\Omega_{\rm DM,0} &=&
\frac{\,\rho_{\Phi,0}^{}\,}{\rho_{c,0}^{}}
= \frac{\,m_{\Phi}^{}n_{\Phi,{\rm osc}}^{}\,}{\rho_{c,0}^{}s_{{\rm osc}}^{}}s_0^{}\,,
\quad\Longrightarrow
\\
m_{\Phi}^{} &\simeq&
\xi_{{\rm DM}}^{2}(3\!\times\!\!10^{-5})\!
\(\!\frac{g_{*}^{}}{\,106.75\,}\!\)^{\!\!1/2}\!\!
\(\!\frac{\,\phi_{{\rm inf}}^{}\,}{\,10^{13}{\rm GeV}\,}\!\)^{\!\!\!-4}{\rm eV}\,,
\label{eq:edensity}
\eeqa
\eeqs
%
%we require the fraction $\xi_{{\rm DM}}^{}$ to be less than 1\%,
where $n_{\Phi}^{}$ is the number density of the scalar particle, $s$ is the entropy density,
and $0$ (osc) in the subscript indicates that the quantities are evaluated
for today ($t_{{\rm osc}}^{}$). For our current analysis, we will require the DM fraction
from boson stars to be 1\%, $\,\xi_{{\rm DM}}^{}\simeq 0.01\,$,\,
which serves as a benchmark condition.
As explained in Appendix\,\ref{AppendA}, for
$\,m_{\Phi,{\rm inf}}\simeq H_{{\rm inf}}^{}$,\,
the CMB constraint on the isocurvature parameter $\,\beta_{{\rm iso}}^{}\!\!<\!0.038\,$
at $\,k_{{\rm mid}}^{}\!=\!0.050{\rm Mpc^{-1}}$\,
leads to the condition
$\,\phi_{{\rm inf}}^{}\!\lesssim\! 0.25H_{{\rm inf}}^{}$\,.\,
Using $\,H_{{\rm inf}}^{}\simeq 6.3\!\times\!10^{13}\sqrt{r/0.064\,}{\rm GeV}$
\cite{Bertolami:2016ywc} and the CMB constraint on the tensor-to-scalar ratio $\,r\leqq 0.064\,$ \cite{Akrami:2018odb}, we finally derive
\beqa
m_{\Phi}~\gtrsim~
\xi_{{\rm DM}}^{2}\!\times\!4.875\!\times\!10^{-6}
\!\times\!\!\(\!\frac{g_{*}}{\,106.75\,}\!\)\!\!\times\!\!
\(\!\frac{r}{\,0.064\,}\!\)^{\!\!-2}{\rm eV} .
\label{eq:massTSratio}
\eeqa

For $\,10.75\!\leqq\!g_{*}^{}\!\leqq\!106.75$,\footnote{To be consistent with the MACHO constraint on the scalar particle mass $m_{\Phi}^{}$,\,
we have $\,m_{\Phi}^{}\!\gtrsim\! \mathcal{O}(10^{-13})$eV.
The effective number of relativistic degrees of freedom
$\,g_{*}^{}$ would be $\,g_{*}^{}(T\!\simeq\! 10{\rm MeV})\sim 10.75$\,
at the time when $\,H\simeq m_{\Phi}^{}\!\sim\!\mathcal{O}(10^{-13}){\rm eV}$ holds.}
we may expect the allowed minimum scalar mass for the boson stars with
$\,\xi_{{\rm DM}}^{}\!=\!10^{-2}$ (or $\xi_{{\rm DM}}^{}\!=\!10^{-1}$) to be
$\,m_\Phi^{}\!\gtrsim\! \mathcal{O}(10^{-10})$eV
(or $\,m_\Phi^{}\!\gtrsim\!\mathcal{O}(10^{-8}){\rm eV}$).
In comparison with the MACHO constraints in Table\,\ref{table:phimass},
we find that the fraction $\xi_{{\rm DM}}^{}=0.01$ is consistent with the current
cosmological and astrophysical data for the scalar mass-range
$\,\mathcal{O}(10^{-10}){\rm eV}< m_{\Phi}^{} <\mathcal{O}(10^{-3}){\rm eV}$,\,
while the fraction $\,\xi_{{\rm DM}}^{}=0.1$\,
is excluded because the MACHO and CMB constraints on $m_{\Phi}^{}$ do not overlap.
Hence, in the following analyses, we focus on the benchmark case where the boson stars
are responsible for 1\% of the DM population in the Universe with the mass range
$\mathcal{O}(10^{-10}){\rm eV}\! < m_{\Phi}^{}\! < \mathcal{O}(10^{-3}){\rm eV}$.

\vspace*{2mm}
\subsection{Probe by Lensing of Fast Radio Bursts}
\label{sec:FRB}
\label{sec:4.2}
\vspace*{2mm}

The origin of the fast radio bursts (FRBs)
remains unknown, and we expect that the boson stars can cause
some portion of the generated FRBs to be lensed by their own gravitational field.
Probing the dark matter (DM) mass and the fraction of DM $\xi_{\rm DM}^{}$
as occupied by the primordial black hole was studied before\,\cite{Munoz:2016tmg}.
It was also extended to the applications for probing the parameter space
of exotic compact objects including the mini-boson stars,
the boson stars with quartic self-interaction, and the fermion stars\,\cite{Laha:2018zav}.
In this subsection, following \cite{Munoz:2016tmg,Laha:2018zav},
we study how to use the lensing of FRBs to probe the presence of boson stars
as well as to constrain the parameter space $(m_{\Phi}^{},\,\Lambda)$
of our boson star models. The FRBs and their usage were discussed before\,\cite{Munoz:2016tmg,Laha:2018zav,Ravi:2019acz}.
For the purpose of the present study, we briefly explain the computation
of the optical depth in Appendix\,\ref{AppendB}.

\vspace*{1mm}

In our analysis, we assume the total number of observable FRB signals in the near future to be $\,N_{{\rm FRB}}^{}=10^{4}$\, according to \cite{Amiri:2018qsq}
where it was shown that CHIME could detect thousands of FRBs throughout
its envisioned project lifetime (3\,years).
For a set of values of the DM fraction $\xi_{{\rm DM}}^{}$ as contributed by the boson stars,
the boson star mass $M_{L}^{}$ as a lens, and the time delay $\overline{\Delta t}$\, between
the two images from the FRB gravitational lensing, we can estimate the integrated-optical
depth $\,\overline{\tau}$\, according to the procedure shown in Appendix\,\ref{AppendB}.
Here $\,\overline{\tau}$\, represents the probability for a FRB to be lensed
by the presence of a compact object.
We find that requiring $\,\overline{\tau}\geqq 10^{-4}$\,
can make a part of the parameter space of the boson star model be probed by detecting the FRB lensing signals.

\begin{figure}[t]
\centering
\includegraphics[width=7.8cm,height=6.0cm]{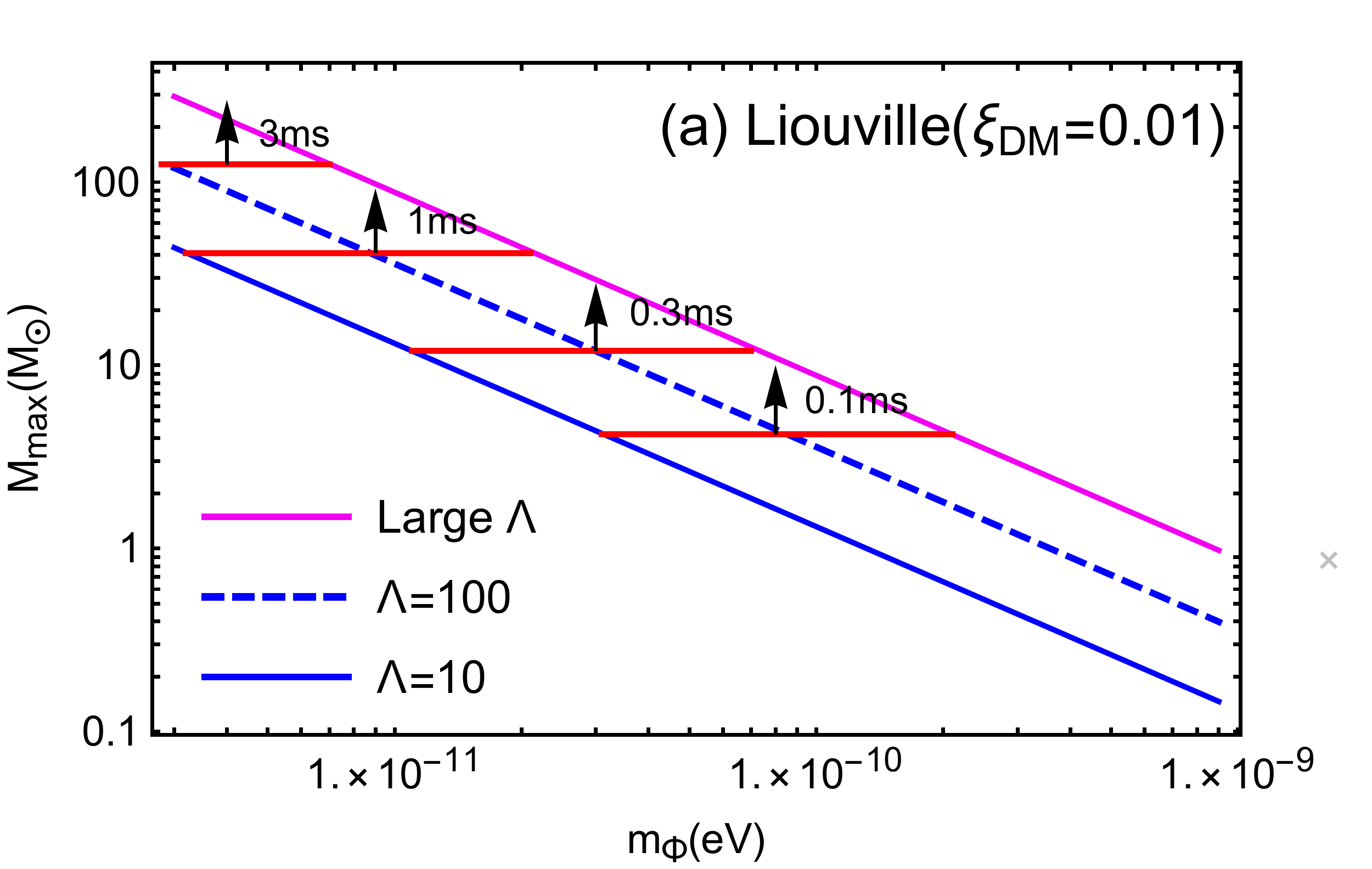}
%[scale=0.308]{figs/Figure10a.pdf}
\hspace*{-5mm}
\includegraphics[width=7.8cm,height=6.0cm]{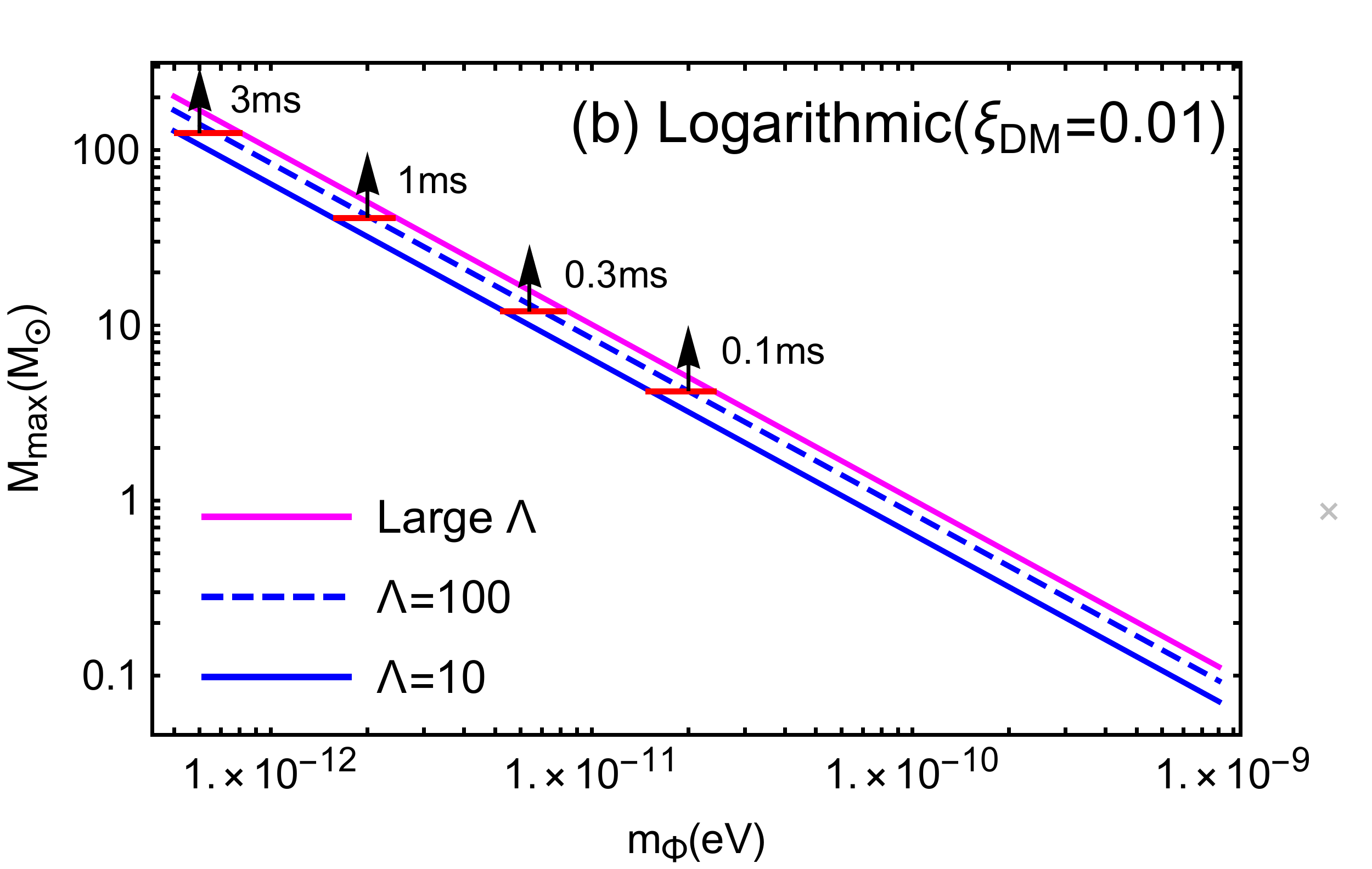}
%[scale=0.308]{figs/Figure10b.pdf}
\caption{Parameter space of the boson star models which can be probed by detecting
the lensed fast radio bursts (FRBs). We assume that the DM fraction contributed
by boson stars is $\,\xi_{{\rm DM}}^{}\!=\!1\%$\,.\,
In each plot, the regions above the red horizontal lines can be probed by the lensed FRBs
with the corresponding time delays $\overline{\Delta t}$\,.\,
For each given coupling strength parameter, the blue and magenta lines show the relation
between the scalar particle mass and the maximum mass of the boson star.}
\label{fig:FRB}
\label{fig:10}
\end{figure}

\vspace*{1mm}

We present in Fig.\,\ref{fig:FRB} the parameter space of the two benchmark models of boson stars that can be probed
by detecting the lensed FRB signals by the boson star.
We consider four different reference times
(\,$\overline{\Delta t}=0.1,\, 0.3,\, 1,\, 3$\,ms)
following \cite{Munoz:2016tmg,Laha:2018zav} and an exemplary DM fraction contributed by boson stars (\,$\xi_{{\rm DM}}^{}\!=\!0.01$).
As explained in \cite{Laha:2018zav}, FRB170827 shows the temporal profile with
three different components. The narrowest component is characterized
by the width $\sim\!30{\rm\mu s}$,\,
which justifies the choice of the smallest reference time, i.e., 0.1ms if one intends to detect the lensed image of mini-bursts
(cf.\ Fig.\,2 of \cite{Farah:2018buz}).
The solid and dashed blue lines in Fig.\,\ref{fig:FRB} represent the relation
between the scalar particle mass $m_{\Phi}^{}$ and the maximum achievable
boson star mass $M_{\max}^{}$ for the dimensionless coupling strength
$\,\Lambda\!=\!10$ and $\,\Lambda\!=\!100$, respectively.
The relation line with the coupling strength between $\,\Lambda\!=\!10$ and $\,\Lambda\!=\!100$ is located
between the solid and dashed blue lines. The relation lines are plotted based on our results in Sec.\,\ref{sec:numerical}.Besides, the magenta line is plotted with
the asymptote of the maximum boson star mass when $\Lambda$ approaches infinity,
which we inferred by fitting and extrapolating $M_{{\max}}^{}$ for $\,\Lambda\leqq 100\,$ as numerically obtained in Sec.\,\ref{sec:numerical}.
The horizontal red lines denote the minimum boson star mass above which
detecting the FRBs lensed by the boson star with the corresponding time delays $\overline{\Delta t}\,$ becomes possible.
These red lines are derived by requiring the probability
$\overline{\tau}\!\geqq\! 10^{-4}$
for the benchmark DM fraction $\,\xi_{{\rm DM}}^{}\!=\!0.01$\,.

\vspace*{1mm}

The optical depth computation reveals that the smaller time delay
between the two images can better probe the presence of the boson star
with smaller mass. As discussed in our previous Sec.\,\ref{sec:scalarmass},
the allowed mass range of the scalar particles which compose the boson star reads $\,\mathcal{O}(10^{-10}){\rm eV}\!\!<\!m_{\Phi}^{}\!\!<\!\mathcal{O}(10^{-3}){\rm eV}$.
Hence, what matters is whether $M_{{\max}}^{}$
resulting from the scalar mass range of $m_{\Phi}^{}$ for a given potential
can be large enough to be probed by detecting the lensed FRB signals.
For the Liouville potential and in the scalar mass range
$\,\mathcal{O}(10^{-10}){\rm eV}\!\!<\!m_{\Phi}^{}\!\!<\!\mathcal{O}(10^{-3}){\rm eV}$,\,
we find that $\,M_{{\max}}^{}$\, with $\,\Lambda\gtrsim 100\,$ becomes large enough
to be probed by FRB lensing with $\overline{\Delta t}=0.1$ms.
In contrast, the logarithmic potential is unable to produce a large enough $M_{{\max}}^{}$ to be probed by FRB lensing in the same scalar mass range
$\,\mathcal{O}(10^{-10}){\rm eV}<m_{\Phi}^{}<\mathcal{O}(10^{-3}){\rm eV}$.\footnote{%
We note that a smaller DM fraction $\,\xi_{\rm DM}^{}\!< 0.01$
can relax the lower bound on $\,m_\Phi^{}\,$ according to Eq.\eqref{eq:massTSratio}.
Then, our Fig.\ref{fig:10}(b) shows that for a low enough scalar mass $m_{\Phi}^{}$ to produce
the total boson star mass $\,M\!\!\gtrsim \!5M_{\odot}^{}$,\, this is within $m_{\Phi}^{}$ range
as allowed by the current $\beta_{\rm iso}^{}$ constraint.
So, if the planned future CHIME experiment allows
the use of FRB lensing with $\overline{\Delta t\,}=0.1{\rm ms}$,\,
then the lensed FRB signals can be detected
by CHIME even for the Logarithmic potential.}
For a given scalar particle mass $m_{\Phi}^{}$,
the logarithmic potential tends to produce a boson star with lighter total mass than that of the Liouville potential.
Hence, it is more challenging to probe the boson stars with logarithmic potential
by FRB lensing because it requires a shorter time delay between the two images.

\vspace*{2mm}
\subsection{Probe by Gravitational Wave Detection}
\label{sec:GW}
\label{sec:4.3}

If the boson stars described by our scalar potential models appear in the Universe,
we may expect the two neighboring boson stars to form a binary system,
go through inspiraling motion, and then merge to produce gravitational waves (GWs).
The frequency of the GW depends on the mass and compactness of the boson stars. Hence they could be invoked to infer the information about the the scalar particle mass
$m_{\Phi}^{}$ and the coupling strength $\Lambda$ of our potential models introduced in Sec.\,\ref{sec:potential}. Using the LIGO GW measurement to probe the mini-boson star or boson stars with a
quartic potential was discussed before\,\cite{Giudice:2016zpa}.

\vspace*{1mm}

In this subsection, we apply our two benchmark potential models
(Sec.\,\ref{sec:potential}) to study
how the scalar particle mass $m_{\Phi}^{}$ and coupling strength $\Lambda$
can be possibly probed by the LIGO GW detection.
In Appendix\,\ref{AppendC}, we will explain further about
obtaining the sub-region in the plane of $(C,\,M_{\max}^{})$
which corresponds to the high enough signal to noise ratio (SNR)
for the LIGO GW detector.
Here we assume a boson star accomplishes its allowed maximum mass $M_{\max}^{}$
and compactness $C_{\max}^{}$ for stability.
For the simplicity of analysis, we consider that the two merging boson stars share
the common mass and compactness. We shall address the following questions:
for what values of $(m_{\Phi}^{},\,\Lambda)$, can the merger of two boson stars
produce GW signals whose frequency falls into the sensitive range of the LIGO detector?
Could such $m_{\Phi}^{}$ values be consistent with the mass range as we discussed in Sec.\,\ref{sec:scalarmass}?

\begin{figure}[t]
\centering
\includegraphics[scale=0.36]{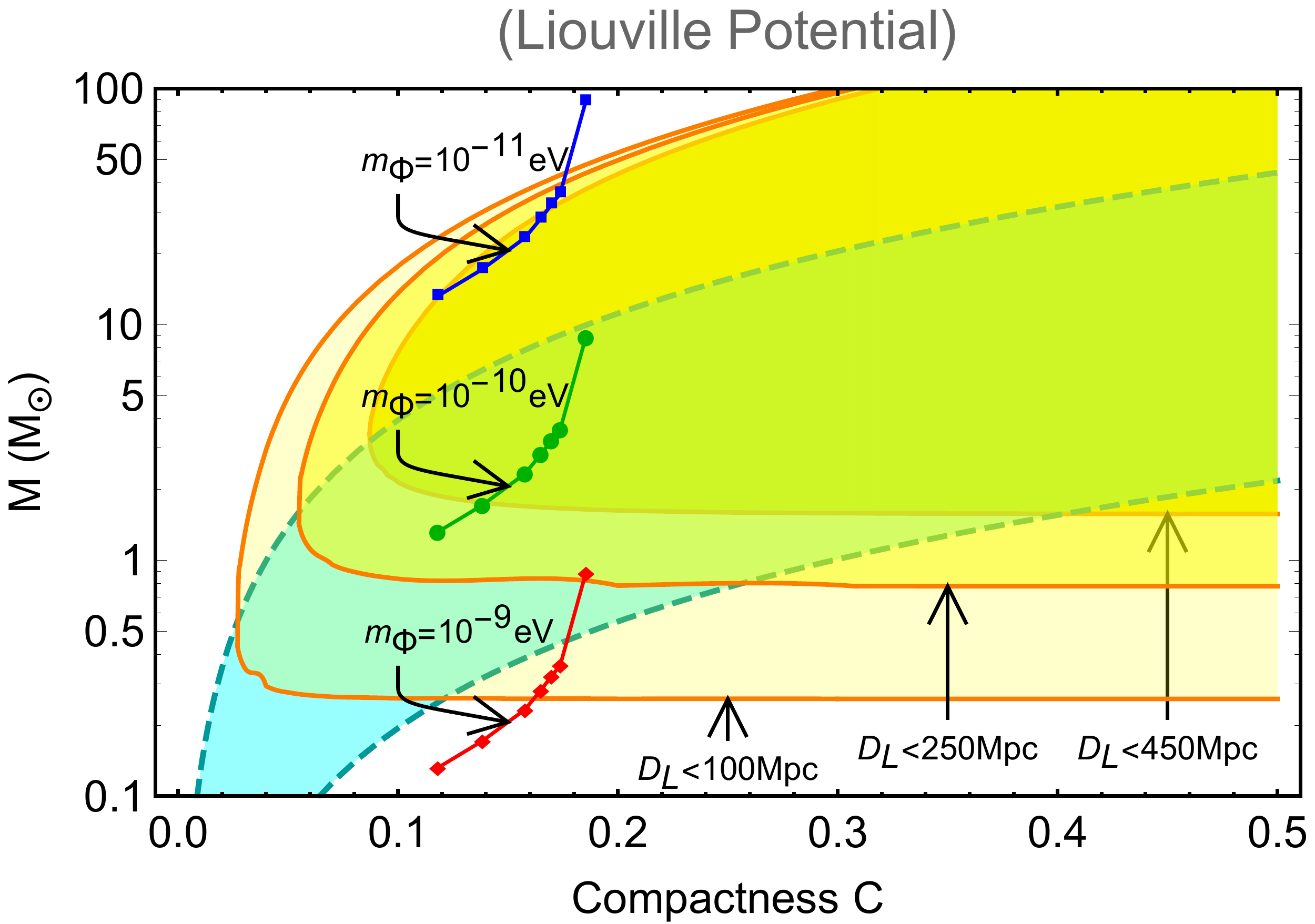}
\\[1mm]
\includegraphics[scale=0.36]{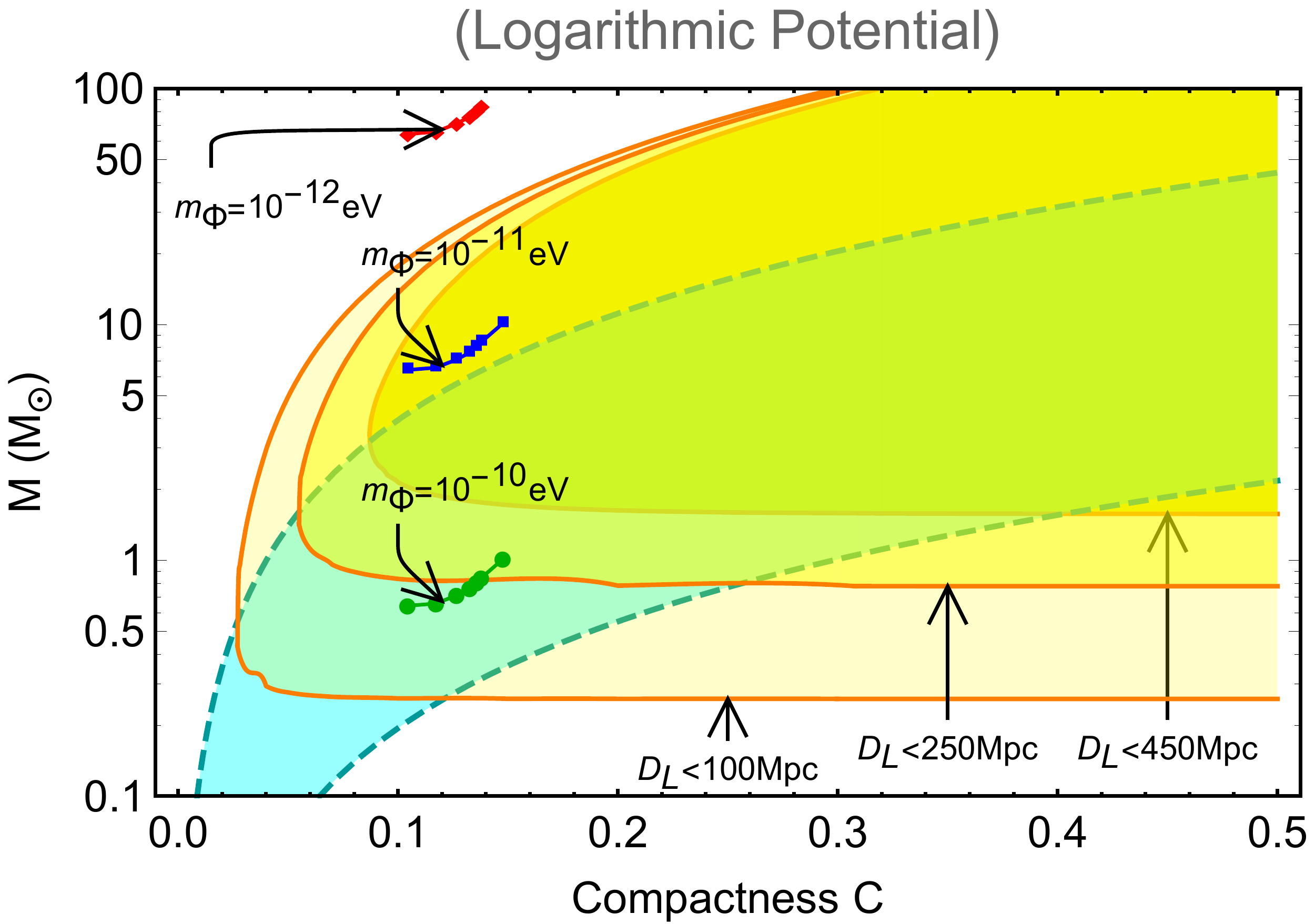}
\vspace*{-1.5mm}
\caption{The boson star mass $M$ (in unit of solar mass) versus its compactness $C$.
Colored regions show the parameter space where the GW signals from the merger of binary boson stars
can be probed by LIGO. The overlapping sub-region between the cyan and yellow regions
ensures high enough signal to noise ratio for the LIGO detection.
The (red, green and blue) dots correspond to the $(C_{\max}^{},\,M_{\max}^{})$ values of the
boson stars with sample scalar mass $m_{\Phi}^{}$ and self-interaction strength $\Lambda$\,.\,
}
\label{fig:GW}
\label{fig:11}
\vspace*{-3mm}
\end{figure}

\vspace*{1mm}

We present the findings from our analysis in Fig.\,\ref{fig:GW}.
For a set of values of $(C_{\max}^{},\,M_{\max}^{})$ within the sub-region with cyan color,
the merger of the binary boson stars will generate GW signals whose frequency
lies in the range ($50{\rm Hz}\leqq \nu\leqq 1000{\rm Hz}$)
where the noise spectral density $S_{n}^{}(\nu)$\,
is low enough for the LIGO sensitivity.
In addition, the GW signals from the boson stars within the sub-region
of yellow color satisfy the high enough SNR with $\,\rho\!\ge\!8\,$,\,
where $\rho$ is defined in Eq.(\ref{eq:rho}).
The (red, green and blue) dots correspond to the $(C_{\max}^{},\,M_{\max}^{})$ values
of the boson stars formed by the scalar particles of mass $m_{\Phi}^{}$
and self-interaction coupling strength $\Lambda=10-100$. For each color, we also show the asymptote values of $(C_{\max}^{},\,M_{\max}^{})$ for a large $\Lambda$ which are obtained by fitting and extrapolating the results in Sec.\,\ref{sec:numerical}.\,
The higher coupling strength $\Lambda$ leads to the higher $C_{\max}^{}$ and $M_{\max}^{}$.

\vspace*{1mm}

For the case of Liouville potential, we find that boson stars
with scalar particle mass $\,m_{\Phi}^{}\!=\!10^{-10}$eV (for any $\Lambda$)\, or $\,m_{\Phi}^{}\!=\!10^{-9}$eV (for $\Lambda>100$) are lying in the overlapping region between the cyan and yellow areas,
so their merger will produce GW signals whose frequency can probed by the LIGO detector.
For $\,m_{\Phi}^{}\!=\!10^{-10}$eV, if the self-interaction strength $\Lambda\leqq 20$,
the boson star merger at the luminosity distance $D_L^{}\!\!<\!250{\rm Mpc}$ may produce detectable GWs by LIGO. For $\,m_{\Phi}^{}\!=\!10^{-9}$eV, if $\Lambda>100$, then those mergers that occurred at the luminosity distance $D_L^{}\!\!<\!100{\rm Mpc}$ may produce detectable GWs. For boson stars with logarithmic potential, the scalar particle mass also needs
to be $\,m_{\Phi}^{}=10^{-10}$eV so that it can lie in the overlapping region
of the $(C_{\max}^{},\,M_{\max}^{})$ plane.
We find that even the case with strong self-interaction strength $\,\Lambda>100$\,
requires the merger of boson stars to occur at $\,D_L^{}\!<250\,\text{Mpc}$ for the LIGO detection.\,
In comparison, for the same scalar particle mass $m_{\Phi}^{}$ and coupling strength $\Lambda$,\,
the Liouville potential achieves a higher compactness $C_{\max}^{}$ and
boson star mass $M_{\max}^{}$. Hence, the farther merging event can be probed
for boson stars with Liouville potential rather than the logarithmic potential.
In addition, the mass $\,m_{\Phi}^{}\!=\!10^{-10}$eV\,
which can be probed by the GW detection lies in the range
$\mathcal{O}(10^{-10}){\rm eV}\!\lesssim\! m_{\Phi}^{}\!\lesssim\!\mathcal{O}(10^{-3}){\rm eV}$
for both potentials.
This implies that GW detection can compensate the FRB lensing probe
for searching boson stars with the logarithmic potential.

%\newpage
\section{Conclusions}

In this work, we studied the dynamics of boson stars that consist of a new type of light
singlet scalar particles with nontrivial self-interactions. Such boson stars may
compose a small fraction of the dark matter in the Universe.
We considered such light scalars with two distinctive self-interaction potentials as the benchmarks ---  the Liouville potential (which generates an
infinite series of repulsive forces)
and the logarithmic potential (which generates an infinite series of pairs of
attractive and repulsive forces with a net attractive force).

\vspace*{1mm}

In Sec.\,\ref{BSmodel}, we introduced the Einstein-Klein-Gordon equations
with which we can study properties of the scalar field $\phi(r)$ in the curved spacetime background.
We discussed the underlying physics to motivate the two benchmark scalar potentials. We further defined the total mass and compactness of the boson star.

\vspace*{1mm}

In Sec.\,\ref{sec:results}, we presented a numerical approach to
solve the Einstein-Klein-Gordon equations and then applied a semi-analytic approach.
We focused on the scalar field configuration for the ground state of the boson star
which has the global maximum at the origin and approaches zero at the spatial infinity
without a node. We found that the total mass of the stable boson star continues
to increase with the rise of $\phi_{0}^{}$ value until $\phi_{0}^{}$ reaches
a critical value $\,\phi_{0}^{\star}$\, that corresponds to the maximum total mass, where $\,\phi_0^{}=\phi(r)|_{r=0}^{}$\,.\,
Beyond $\phi_{0}^{\star}$\,,\, the boson star becomes unstable.
Furthermore, the stable boson star with either scalar potential was found to have
negative binding energies. Intriguingly, as self-interaction strength increases,
the negative binding energy for a boson star with the logarithmic potential decreases
in magnitude up to the coupling strength $\Lambda\!\sim\!5$ and then increases beyond this coupling value\,;\,
while for the case of Liouville potential it monotonically increases in magnitude.
In addition, we studied the maximum compactness as a function of the coupling strength $\Lambda$,
and obtained important insights.
For the boson star with Liouville potential, the maximum compactness can reach as high as
$\,C_{\max}^{}\!\sim\!0.18$\, for large coupling, and is larger than the case with
a repulsive quartic interaction or a logarithmic interaction.
On the other hand, the case of the logarithmic potential showed a slight deficit as compared
to the case of a repulsive quartic potential.
In the last part of this section, we applied the Swampland conjecture
and found that the maximum compactness $C_{\max}^{}$
obtained by the full numerical computation for both potentials arises
from low energy effective scalar field theories which could be
UV-completed by a consistent quantum gravity theory.

\vspace*{1mm}

In Sec.\,\ref{sec:AP}, we studied the lensing of FRBs and the GW detection of LIGO
as two astrophysical methods to probe the presence of boson stars and the
parameter space of their corresponding scalar potentials.
Given the future planned CHIME FRB detection experiment, we expect that boson stars
with a total mass as light as $M\!=\!\mathcal{O}(1)M_{\odot}^{}$
can be probed. Furthermore, LIGO will be sensitive to the GW signals from merging boson stars
with a total mass $\,M\!=\!\mathcal{O}(1)M_{\odot}^{}$\,.\,
As for the individual scalar particle mass, LIGO is most sensitive to
$\,m_{\Phi}^{}\!\simeq\!10^{-10}{\rm eV}$ for both potentials,
whereas the FRB lensing detection can probe a sizable mass range $\,m_{\Phi}^{}\!\simeq\!\mathcal{O}(10^{-12}\!-\!10^{-10}){\rm eV}$,\,
depending on the time delay between two lensed FRB images by boson stars.
We demonstrated that the current constraints on the DM fraction
as contributed by MACHO and the CMB constraint on the cold DM isocurvature modes
give the allowed scalar mass range
$\,\mathcal{O}(10^{-10}){\rm eV}\!\!<\!m_{\Phi}^{}\!\!<\!\mathcal{O}(10^{-3}){\rm eV}$,\,
where the scalar particles ($\Phi$) compose the boson stars.
To suppress non-thermal production of the scalar particles in the early Universe,
we considered a scenario where the scalar field has gravitationally induced
coupling to the inflaton during the inflation.
Because of this coupling, the effective mass of the scalar particle during inflation
is comparable to the Hubble expansion rate at that time. After inflation ends, the scalar mass reduces to its original range
$\,\mathcal{O}(10^{-10}){\rm eV}\!<\! m_{\Phi}^{}\!<\! \mathcal{O}(10^{-3}){\rm eV}$.
Applying this mass range, we found that searching the boson stars
with Liouville potential could be probed by both FRB lensing and GW detection. For boson stars with logarithmic potential,
we find that the LIGO GW detection can probe the presence of the boson stars
and the parameter space of the corresponding scalar theory.
But, the FRB lensing can hardly probe boson stars with logarithmic potential
since the mass range $\mathcal{O}(10^{-10}){\rm eV}\!<\!m_{\Phi}^{}\!<\!\mathcal{O}(10^{-3}){\rm eV}$
does not produce a large enough total mass of boson star as needed by a valid
FRB lensing.
We anticipate the synergy between the FRB lensing and GW detection can help probing
the boson stars, especially for the scalar particle mass around
$\,m_{\Phi}^{}\simeq\mathcal{O}(10^{-10}){\rm eV}$.

\vspace*{8mm}
\noindent
{\bf\large Acknowledgments}
\\[1mm]
We thank Mark Hertzberg for discussing boson stars. GC thanks Ranjan Laha for a useful discussion about the lensing of a fast radio burst, and Tsutomu Yanagida and Yue Zhao for a helpful discussion. This research was supported in part
by the National Key R\,\&\,D Program of China (No.\,2017YFA0402204),
by the National NSF of China (under grants 11275101, 11835005),
by the TDLI and SJTU Postdoctoral Fellowship grants,
by the Shanghai Laboratory for Particle Physics and Cosmology (No.\,11DZ2260700),
by the Office of Science and Technology, Shanghai Municipal Government
(No.\,16DZ2260200), and by the CAS Center for Excellence in Particle Physics (CCEPP).

\appendix

\vspace*{10mm}
\noindent
{\bf\Large Appendix}

\vspace*{1mm}
\section{Initial Scalar Displacement after Inflation}
\label{AppendA}
\label{app:A}

In this Appendix, we follow the logic of \cite{Riotto:2002yw,Bertolami:2016ywc}
to obtain the expression of the scalar field displacement from the global minimum of the potential
at the end of the inflation ($|\Phi_{{\rm inf}}^{}|=\phi_{{\rm inf}}^{}$)
in terms of the tensor-to-scalar perturbation ratio $r$\,.\,
For a massive scalar field with  $\,m_{\Phi}^{}<1.5H_{{\rm inf}}^{}$,\,
the Fourier mode of the field fluctuation on the super-horizon scale
is given by\,\cite{Riotto:2002yw},
\beqa
|\delta\phi_{k}^{}|\,\simeq\,
\frac{H_{{\rm inf}}^{}}{\,\sqrt{2k^{3}\,}\,}
\(\!\frac{k}{\,aH_{{\rm inf}}^{}\,}\!\)^{\!\!\!\frac{3}{2}-\nu_{\!\phi}^{}}\,,
\label{eq:fluctuation}
\eeqa
where $\nu_{\phi}^{}=\sqrt{(9/4)\!-\!(m_{\Phi}/H_{{\rm inf}})^{2}\,}$.
Integrating over all the super-horizon modes,
one obtains the variance of the field value\,\cite{Bertolami:2016ywc},
\beqa
\left<\phi^{2}\right> \,\simeq\,
\frac{1}{\,3\!-\!2\nu_{\Phi}^{}\,}
\(\!\frac{\,H_{\rm inf}^{}\,}{2\pi}\!\)^{\!\!2}.
\label{eq:variance}
\eeqa
This result can be used for $\phi_{{\rm inf}}^{}$ value by approximating
$\,\phi_{{\rm inf}}^{}\!\simeq\sqrt{\left<\phi^{2}\right>\,}$.\,
Note that $\phi_{{\rm inf}}^{}$ will diverge for $m_{\Phi}^{}\ll H_{{\rm inf}}^{}$,\,
so it would generate too much scalar dark matter abundance.
Hence, it is essential for our study to have $m_{\Phi}^{}\simeq H_{{\rm inf}}^{}$ during inflation which could be realized by gravitationally induced coupling between the scalar and inflaton \cite{Bertolami:2016ywc}. From Eqs.(\ref{eq:fluctuation})-(\ref{eq:variance}), we see that
the isocurvature power spectrum is obtained as
\beqa
\Delta_{I}^{2}\,= \(\!\frac{2}{\,\phi_{{\rm inf}}\,}\!\)^{\!\!2}
P_{\delta\phi_{k}}^{} = (3\!-\!2\nu_{\Phi}^{})\!\!
\(\!\frac{k}{\,aH_{{\rm inf}}^{}}\!\)^{\!\!3-2\nu_{\Phi}}\,,
\label{eq:isoPS}
\eeqa
where $\,P_{\delta\phi_{k}}\!\!=\!(k^{3}/2\pi^{2})|\delta\phi_k^{}|^{2}$
\cite{Riotto:2002yw}. The measurement on the isocurvature mode is written in terms of this isocurvature power spectrum and the amplitude of the adiabatic curvature power spectrum
($\Delta_{R}^{2}(k)$),\,
\beqa
\beta_{{\rm iso}}^{}(k) \,=\,
\frac{\Delta_{I}^{2}(k)}{\,\Delta_{\mathcal{R}}^{2}(k)\!+\!\Delta_{I}^{2}(k)\,}\,.
\label{eq:isoPS}
\eeqa
Since $\delta\phi$ and $\delta\chi$ are independent,
the CDM isocurvature modes are uncorrelated with the adiabatic modes \cite{Bertolami:2016ywc}.
Hence, combined with
$\,\Delta_{R}^{2}(k_{{\rm mid}}^{}) \simeq 2.1\!\times\!10^{-9}$ \cite{Akrami:2018odb},\,
we find that the constraint on $\,\beta_{{\rm iso}}\!<\!0.038$\,
at $\,k_{{\rm mid}}^{}\!=0.050{\rm Mpc^{-1}}$ \cite{Akrami:2018odb}
leads to $\,\nu_{\Phi}^{}\lesssim1.297$\, for 55 e-folds of inflation,
and thus $\,\phi_{{\rm inf}}^{}\lesssim 0.25H_{{\rm inf}}^{}$
via Eq.(\ref{eq:variance}). We have used this result to derive the lower bound on the scalar particle mass $m_{\Phi}^{}$ in Eq.(\ref{eq:massTSratio}).

\vspace*{1mm}
\section{Computation of the Optical Depth}
\label{AppendB}
\label{app:B}

For presenting our analysis in Sec.\,\ref{sec:4.1}, following \cite{Munoz:2016tmg,Laha:2018zav},
we explain the procedure of computing the integrated optical depth
$\,\bar{\tau}(M_{L}^{})$\, in this Appendix.

\vspace*{1mm}

The observable $\,\bar{\tau}(M_{L}^{})$\, is interpreted as the probability for
a FRB lensed by a compact object with mass $M_{L}^{}$.\,
The time delay between the two images resulting from the lensing of FRB
by a compact object of mass $M_{L}^{}$ and location $z_{L}^{}$ is given as follows,
\beqa
\Delta t\,=\,\frac{\,4\GN M_L^{}\,}{c^{3}}(1\!+\!z_{L}^{})
\!\!\left[\frac{\,y}{2}\sqrt{y^{2}\!+\!4\,} +
\log\!\(\!\frac{\sqrt{y^{2}\!+\!4\,}\!+\!y\,}{\sqrt{y^{2}\!+\!4\,}\!-\!y\,}\!\)\!\right]\!,
\label{eq:delayT}
\eeqa
where $y=\beta/\theta_{E}^{}$.\,
The parameter $\beta$ is the angular impact parameter, and the angle $\theta_{E}^{}$
is an angular Einstein radius determined by the angular diameter distances
to the source of FRBs ($D_{s}^{}$) to a lens ($D_{L}^{}$),
and between the two ($D_{LS}^{}$), as given by
\beqa
\theta_{E}^{}=\,
2\sqrt{\frac{\GN M_L^{}}{c^{2}}\frac{D_{LS}^{}}{D_{S}^{}D_{L}^{}}}\,.
\eeqa

The following conditions to insure the strong enough lensing of a FRB give us $(y_{\min}^{},\,y_{\max}^{})$.\,
The parameter $R_{f}^{}$ is defined as the ratio of the size of the larger image to that of
the smaller image. Requiring $R_{f}^{}$ to be smaller than the critical value
$\,\overline{R}_{f}^{}=5\,$ provides the following maximal allowed value of $\,y$\,,
\beqa
R_{f}^{}\,=\,
\frac{\,y^{2}\!+\!2\!+\!y\sqrt{y^{2}\!+\!4\,}\,}
     {\,y^{2}\!+\!2\!-\!y\sqrt{y^{2}\!+\!4}\,}
\leqq \overline{R}_{f}^{} = 5 \,, \quad\To\quad
y_{\max}^{} =\!\(\!\frac{1\!+\!\overline{R}_{f}^{}}
{\sqrt{\overline{R}_{f}^{}}}-2\!\)^{\!\!\!\!\fr{1}{2}}\!\!
\simeq\, 0.8266\,.~~~~~
\eeqa
\noindent
Demanding that the time delay (\ref{eq:delayT}) be greater than a reference time
$\,\overline{\Delta t}$\, gives the minimum value
$y_{\min}^{}$.\footnote{
For the current numerical estimate of $y_{{\min}}^{}$,\,
we expand the expression in the square bracket of Eq.(\ref{eq:delayT})
and use the approximation,
\beqa
\Delta t\simeq\frac{\,8\GN M_L^{}\,}{c^{3}}\(1+z_L^{}\!\)y\,.
\eeqa}
For a given set of $(y_{\max}^{},\,y_{\min}^{})$,
the optical depth $\tau$ as a function of the mass of the lens $M_{L}^{}$
and the source position $\,z_{S}^{}\,$ can be computed from
\beqa
\tau(M_L^{},z_S^{}) \,=\, \frac{3}{2}\xi_{{\rm DM}}^{}
\Omega_c^{}\!\int_{0}^{z_{S}^{}}\!\!\!\dd z_{L}^{}\,\frac{H^{2}_{0}}{\,cH\!(z_{L})\,}
\frac{D_L^{}\!D_{LS}^{}}{D_S^{}}(1\!+\!z_L^{})^{2}\!
\left[y^2_{\max}-y^2_{\min}(M_L^{},z_L^{})\right],
\hspace*{10mm}
\label{eq:tau}
\eeqa
where $H$ is the Hubble expansion rate and $\,\xi_{{\rm DM}}^{}\,$ is the fraction of DM
provided by the compact objects which cause the lensing of FRBs.
We choose the values of cosmological parameters for computing $\tau$
from Ref.\,\cite{Aghanim:2018eyx}.
Different choices of $\,\overline{\Delta t}\,$
and $M_{L}^{}$ would lead to different
$y_{{\min}}^{}$ and thus different $\tau$.\,
For the present study, we assume the constant density redshift distribution function
for sources\,\cite{Oppermann:2016mzk},
\beqa
N_{{\rm const}}(z)\,=\,
\mathcal{N}_{{\rm const}}^{}
\frac{\chi^{2}(z)}{\,H\!(z)(1\!+\!z)\,}e^{-d_{L}^{2}(z)/[2d_{L}^{2}(z_{{\rm cut}}^{})]}\,,
\label{eq:source}
\eeqa
where $d_{L}^{}(z)$ is the luminosity distance,
$\mathcal{N}_{{\rm const}}^{}$ is the nomalization factor,
and $\chi(z)$ is the comoving distance.
We choose $\,z_{{\rm cut}}^{}\!=\!0.5$\,.\,
Finally, convolving $\tau$ in Eq.(\ref{eq:tau}) with the
redshift distribution (\ref{eq:source})
removes $\,z_S^{}$\, dependence and leads to the following integrated-optical depth,
\beqa
\bar{\tau}(M_L^{})=\int \!\!\dd z\,\tau(z,M_L^{})N(z)\,.
\eeqa
This quantity is interpreted as the probability for
a single burst to be lensed.

\section{Sensitive Parameter Space of Boson Stars to LIGO}
\label{AppendC}
\label{app:C}
\vspace*{1.5mm}

In this Appendix, following \cite{Giudice:2016zpa}, we explain the procedure
of determining the parameter space of the physical quantities $(C,\,M_{\max}^{})$
of the boson star which can be probed by the LIGO GW detector. For the simplicity of illustration, we consider the situation where
the two merging boson stars have the same mass and compactness as described by
the same the scalar potential.
The GW emissions from the merger of the binary boson stars are characterized
by the frequency,
\beqa
\nu^{\rm BS} \,=\, \frac{~(C\!/3)^{{3}/{2}}_{}\,}{\,2\pi M\,}\,,
\eeqa
where the parameters $(C,\,M)$ are the common (compactness,\,mass) of the binary boson stars.
Requiring $\nu^{\rm BS}$ to lie within the GW frequency range ($50-1000{\rm Hz}$)
to which LIGO detection is sensitive, one obtains the following relation
\beqa
C^{3/2}\!\times\!6.149M_{\odot}^{}\leqq \,M\leqq C^{3/2}\!\times\!124.451M_{\odot}^{}\,,
\eeqa
which must be satisfied by $(C,\,M_{\max}^{})$ of the binary boson stars to be probed with the low level noise. We draw the corresponding region of the parameter space by the cyan color
as in Fig.\,\ref{fig:GW}. The signal to noise ratio (SNR) of the
GW signals with strain $h(t)$ reads
\beqa
\rho^{2} \,=\, \int_{0}^{\nu^{\rm BS}}\!\!\!\dd \nu\,
\frac{\,\,4|\tilde{h}(\nu)|^{2}\,}{\,S_{n}(\nu)\,}\,,
\label{eq:rho}
\eeqa
where $\,\tilde{h}(\nu)$\, is the Fourier transform of the strain
and $\,S_{n}^{}(\nu)$\, is the noise power spectral density (PSD).
Note that the upper limit of the integral depends on both $C$ and $M$\,.\,
We take $S_{n}^{}(\nu)$ from \cite{GWLigo}. In the quadrupole approximation \cite{Khan:2015jqa}, the strain Fourier transform reads
\beqa
\tilde{h}(\nu)\,\simeq\,\frac{\sqrt{5/24}}{\,\pi^{2/3}D_L^{}\,}
M_{c}^{5/6}(\nu^{\rm BS})^{-7/6}\,,
\eeqa
where $ D_L^{}$ is the luminosity distance for the location at which
the merger of the binary boson stars takes place,
and $M_c^{}$ is the chirp mass defined as
\beqa
M_c^{}\,=\, \frac{(M_{1}^{}M_{2}^{})^{3/5}}{\,(M_{1}^{}\!+\!M_{2}^{})^{1/5}}\,,
\eeqa
with $\,M_{1}^{}\!=\!M_{2}^{}\!=\!M\,$ in our case.
We will consider $D_{L}^{}<450,\,250,\,100$\,Mpc cases
(with the corresponding redshifts $z\ll 1$),
so the redshift effect is negligible as in \cite{Giudice:2016zpa}. Requiring $\,\rho\ge8\,$ for ensuring a large enough SNR, we identify the
yellow colored region in the parameter space of $(C,\,M_{\max}^{})$
in Fig.\,\ref{fig:GW}.

%\vspace*{4mm}

\newpage
\addcontentsline{toc}{section}{References\,}
\bibliographystyle{JHEP}
\bibliography{main}

\end{document}